\documentclass[preprint,showpacs,preprintnumbers,amsmath,amssymb]{revtex4-1}

\usepackage{graphicx}
\usepackage{epstopdf}

\usepackage{dcolumn}
\usepackage{bm}
\usepackage{color}

\usepackage{feynmf}
\usepackage{slashed}
\usepackage{enumerate}

\usepackage[scientific-notation=true]{siunitx} 

\usepackage[utf8]{inputenc}

\usepackage[colorlinks=true,urlcolor=blue,anchorcolor=blue,citecolor=blue,filecolor=blue,linkcolor=blue,menucolor=blue,linktocpage=true]{hyperref} 

\begin{document}

\title{Corrections to electroweak precision observables from mixings of an exotic vector boson in light of the CDF $W$-mass anomaly}

\author{Chengfeng Cai}
\thanks{caichf3@mail.sysu.edu.cn}
\author{Dayun Qiu}
\author{Yi-Lei Tang}
\thanks{tangylei@mail.sysu.edu.cn}
\author{Zhao-Huan Yu}
\thanks{yuzhaoh5@mail.sysu.edu.cn}
\author{Hong-Hao Zhang}
\thanks{zhh98@mail.sysu.edu.cn}
\affiliation{School of Physics, Sun Yat-Sen University, Guangzhou 510275, China}


\begin{abstract}
We enumerate various effective couplings that contribute to the mixings between an exotic vector boson $Z^{\prime}$ and the neutral electroweak vector bosons. The miscellaneous mixing patterns can be evaluated perturbatively. The effective oblique parameters $S^{\prime}$, $T^{\prime}$, and $U^{\prime}$ are calculated to compare with the electroweak precision test results. With the contributions to the non-negligible $U^{\prime}$ parameter from the $\epsilon_{B,W}$ parameters and the aid of some other parameters to cancel the negative $T^{\prime}$, the recent CDF $W$-mass anomaly can therefore be explained.
\end{abstract}


\maketitle
\section{Introduction}

Besides searching for new particles by straightforwardly producing them at the colliders, detecting the tiny deviations of the measured standard model (SM) parameters from their theoretical predicted values is also an important approach to new physics (NP) beyond the SM. In the literature, the Peskin-Takeuchi oblique parameters $S$, $T$, and $U$~\cite{Peskin:1990zt, Peskin:1991sw} are usually applied to test the SM. These parameters are extracted from the self-energy diagrams of the electroweak (EW) vector bosons, and contribute to EW precision observables, such as the $Z$-pole parameters and the masses of the $Z$ and $W$ bosons~\cite{Burgess:1993vc}.  Higher-order  parameters such as $V$, $W$, $X$, and $Y$ are introduced in Ref.~\cite{Barbieri:2004qk}. In particular, when the traditional $S$, $T$, and $U$ parameters are absent, they can dominate the NP contributions to the EW observables.

Recently, the CDF collaboration published a high-precision measurement of the $W$ boson mass $m_W$ based on the $8.8~\si{fb^{-1}}$ of data collected by the CDF~II detector at the Tevatron collider~\cite{CDF:2022hxs},
\begin{equation}
m_W  = 80.4335\pm 0.0094~\si{GeV}.
\end{equation}
This result indicates a $\sim 7\sigma$ deviation from the SM prediction $m_W^\mathrm{SM} = 80.3545\pm 0.0057~\si{GeV}$ given by the global fit of the EW  precision measurements~\cite{deBlas:2021wap}. Such an anomaly had drawn quite a lot of attention and been considered to originate from NP contributions~\cite{Fan:2022dck,Zhu:2022tpr,Lu:2022bgw,Athron:2022qpo,Yuan:2022cpw,Strumia:2022qkt,Yang:2022gvz,deBlas:2022hdk,Du:2022pbp,Tang:2022pxh,Cacciapaglia:2022xih,Blennow:2022yfm,Arias-Aragon:2022ats,Zhu:2022scj,Sakurai:2022hwh,Fan:2022yly,Liu:2022jdq,Lee:2022nqz,Cheng:2022jyi,Song:2022xts,Bagnaschi:2022whn,Paul:2022dds,Bahl:2022xzi,Asadi:2022xiy,DiLuzio:2022xns,Athron:2022isz,Gu:2022htv,Heckman:2022the,Babu:2022pdn,Heo:2022dey,Du:2022brr,Cheung:2022zsb,Crivellin:2022fdf,Endo:2022kiw,Biekotter:2022abc,Balkin:2022glu,Krasnikov:2022xsi,Ahn:2022xeq,Han:2022juu,Zheng:2022irz,Kawamura:2022uft,Peli:2022ybi,Ghoshal:2022vzo,Perez:2022uil,Kanemura:2022ahw,Mondal:2022xdy,Zhang:2022nnh,Borah:2022obi,Chowdhury:2022moc,Arcadi:2022dmt,Cirigliano:2022qdm,Carpenter:2022oyg,Popov:2022ldh,Ghorbani:2022vtv,Du:2022fqv,Bhaskar:2022vgk,Batra:2022org,Cao:2022mif,Zeng:2022lkk,Baek:2022agi,Borah:2022zim,Almeida:2022lcs,Cheng:2022aau,Heeck:2022fvl,Addazi:2022fbj,Lee:2022gyf}.

A wide class of NP effects might contribute to oblique parameters,  and thus shift the $W$ boson mass correspondingly. The EW global fits with the CDF $m_W$ data considered have been performed in Refs.~\cite{Lu:2022bgw, deBlas:2022hdk, Strumia:2022qkt, Bagnaschi:2022whn, Fan:2022yly, Gu:2022htv, Balkin:2022glu, Asadi:2022xiy}. An appropriate loop-level NP contribution implies the corresponding NP scale to be around a few hundred GeV, easily to conflict with current collider bounds, while the tree-level NP scale for interpreting it can be as high as multi-TeV~\cite{Strumia:2022qkt}.

Tree-level corrections to the oblique parameters may come from an exotic neutral vector boson~$Z'$, which naturally appears in many SM extensions, such as grand unified theories~\cite{Robinett:1981yz}, little Higgs models~\cite{Arkani-Hamed:2001nha}, extra dimensions~\cite{Casalbuoni:1999ns}, and a lot of $\mathrm{U}(1)'$ gauge models motivated by various problems~\cite{Leike:1998wr,Langacker:2008yv}. These NP models generally introduce kinetic and mass mixings between the $Z$ and $Z'$ bosons, which contribute to the oblique parameters at tree level~\cite{Holdom:1990xp, Babu:1997st}. Therefore, an exotic $Z'$ boson could be responsible for the CDF $m_W$ anomaly, as discussed in some recent studies~\cite{Strumia:2022qkt, Asadi:2022xiy, Zhang:2022nnh, Du:2022fqv, Zeng:2022lkk, Cheng:2022aau, Alguero:2022est}.  Note that from the effective field theory (EFT) point of view, kinetic mixings between gauge bosons contribute to the $p^4$ terms in the vaccuum polarization amplitudes of the EW gauge fields, leading to higher-order parameters $V$, $W$, $X$, and $Y$. However, combined with the traditional $S$, $T$, and $U$, all these parameters are redundant in fitting the EW precision data at the $Z$-pole, the $W$ mass, and the Fermi constant. Therefore, we can define three effective oblique parameters $S^\prime$, $T^\prime$, and $U^\prime$ which include all order effects and then compare them with the most recent global fit results of $S$, $T$, and $U$ from Ref.~\cite{deBlas:2022hdk}.

In addition to the well-known kinetic mixing between the $\mathrm{U(1)_Y}$ vector boson $B$ and the $Z^{\prime}$ boson, loop-level diagrams might also induce the kinetic mixing terms between the $W^{3}$ and $Z^{\prime}$ bosons as well as various mass mixing terms.  Besides $S'$ and $T'$, a non-negligible $U'$ parameter due to the mixings could also contribute to the $W$ mass, as evaluated in Refs.~\cite{Zeng:2022lkk,Cheng:2022aau} with some specific conditions (Note that the effective $U'$ can originate from higher-order corrections such as $V$, $W$, $X$. See Eq.\eqref{STUprime} and related discussion.). In this paper, a more general case described by an effective Lagrangian is considered.
We will present the perturbative corrections to the $S^\prime$, $T^\prime$, and $U^\prime$ parameters originating from the kinetic and mass mixings between the $Z^{\prime}$ and the SM gauge bosons, in addition to three traditional SMEFT operators. Our strategy of calculating the effective oblique parameters is directly diagonalizing the kinetic and mass matrices. In principle, this procedure automatically includes the corrections from all $(p^2)^i$ orders due to the mixings.

This paper is organized as follows. In Sect.~\ref{Lagrangian}, we briefly introduce the effective Lagrangian for an exotic $Z'$ boson. In Sect.~\ref{ObliqueEvaluation},  the effective oblique parameters $S^\prime$, $T^\prime$, and $U^\prime$ are calculated perturbatively, and the analytical results are presented. In Sect.~\ref{ColliderBounds}, we estimate the constraints for the $Z^\prime$ boson from collider experiments. The numerical results as well as some corresponding discussions have been presented in Sect.~\ref{NumericalResults}, and then we finalize this paper in Sect.~\ref{Summary}.  An EFT analysis is also provided in App.~\ref{Appendix} for cross-checking our calculation.

\section{Effective Lagrangian} \label{Lagrangian}

Besides the SM fields, we introduce an exotic vector field $\hat{Z}^{\prime\mu}$, with kinetic and mass terms given by
\begin{eqnarray}
\mathcal{L} \supset -\frac{1}{4} \hat{Z}^{\prime}_{\mu \nu} \hat{Z}^{\prime \mu \nu} + \frac{1}{2} \hat{m}_{Z^{\prime}}^2 \hat{Z}^{\prime}_{\mu} \hat{Z}^{\prime \mu}, \label{LOriginal}
\end{eqnarray}
where $\hat{Z}^{\prime}_{\mu \nu} \equiv \partial_{\mu} \hat{Z}^{\prime}_{\nu}  - \partial_{\nu} \hat{Z}^{\prime}_{\mu}$.
The $\hat{Z}'$ boson can be either a fundamental gauge boson of an exotic $\mathrm{U}(1)$ gauge group, or a component from a gauge boson multiplet in the framework of a non-abelian gauge group.  The mass term might originate from the vacuum expectation value (VEV) of a Higgs sector, or directly acquired from the Stueckelburg mechanism in the $\mathrm{U}(1)$ case. 

The $\hat{Z}^{\prime}$ boson might mix with the electroweak gauge bosons through the following kinetic mixing terms in the effective Lagrangian,
\begin{eqnarray}
\mathcal{L}_{\rm eff} \supset  &-& \frac{ \epsilon_B}{2} \hat{Z}^{\prime}_{\mu \nu} B^{\mu \nu} -\frac{1}{2 \Lambda_{W}^2 } \hat{Z}^{\prime}_{\mu \nu} W^{a \mu \nu} H^{\dagger} \sigma^a H - \frac{1}{2 \Lambda_{BW}^2} B_{\mu \nu} W^{a \mu \nu} H^{\dagger} \sigma^a H \nonumber \\
&-& \frac{1}{4 \Lambda_{WW}^4}  W^{a \mu \nu} H^{\dagger} \sigma^a H  W^b_{\mu \nu} H^{\dagger} \sigma^b H, \label{LKin}
\end{eqnarray}
where $H$ indicates the SM Higgs doublet, and the $\epsilon_B$ term can be created straightforwardly in the $\mathrm{U}(1)$ case, or can arise together with other $\Lambda_W$, $\Lambda_{BW}$, and $\Lambda_{WW}$ terms through higher-order corrections. The mass terms might also receive corrections, formulated as
\begin{eqnarray}
\mathcal{L}_{\rm eff} &\supset&  \frac{1}{\Lambda_{HD}^2} ( H^{\dagger} D_{\mu} H)^\dag (H^{\dagger} D^{\mu}  H) +  \hat{Z}^{\prime \mu} \left[ i \lambda_{HZ^{\prime}} (D_{\mu} H)^{\dagger} H + \text{H.c.}\right]. \label{LMass}
\end{eqnarray}
All three terms can arise from the loop effects. The $\lambda_{H Z^{\prime}}$ term might originate from something like $\Phi^{\dagger} D^{\prime \mu} \Phi$ where $\Phi$ indicates an exotic Higgs field to break the gauge group corresponding to $\hat{Z}^{\prime}$, or a dummy $v_{Z^{\prime}} e^{i \phi(x)}$ field in the Stueckelberg mechanism. The $\mathrm{U(1)_Y} \times \mathrm{SU(2)_L}$ covariant derivative for the Higgs doublet is  $D_{\mu} = \partial_{\mu} - i \hat{g}^{\prime}_Y B_{\mu}/2 - i \hat{g} \sigma^a W^a_\mu/2$, where the hatted parameters $\hat{g}^{\prime}$ and $\hat{g}$ are the ``original'' coupling constants.

Before proceeding, we would like to make some comments on the possible UV completion of the model. The $S$, $T$, and $U$ parameters corresponds to three SMEFT operators, $B_{\mu\nu}W^{a\mu\nu}H^\dag\sigma^a H$, $|H^\dag D_\mu H|^2$, and $|H^\dag W_{\mu\nu}H|^2$, which can be easily generated by introducing some fermionic~\cite{Cai:2016sjz} or scalar~\cite{Cai:2017wdu} EW multiplets. For example, Ref.~\cite{Cai:2017wdu} had shown that in a dark matter model with a singlet and a doublet scalars, a significant $T$ parameter can be obtained. In particular, when the ratio of the parameters $|\lambda_2/\lambda_3|<0.5$, the $T$ parameter is positive, and thus the $W$ boson mass can be raised. 

The kinetic mixing between $Z'_\mu$ and the SM $B_\mu$ field can be put by hand since it relates to a gauge-invariant renormalizable operator. It can also be generated by loops of some NP fields carring both $\mathrm{U(1)_Y}$ and new $\mathrm{U(1)}$ charges. For the kinetic mixings between $Z'_\mu$ and the SM $W^3_\mu$ field, a simple realization is to generate them by loops of $2n$ scalar EW quadruplet fields, $X_{1,2,...,2n}$, which carry the same new $\mathrm{U(1)}$ charge. Once we choose opposite $\mathrm{U(1)_Y}$ charges for $X_{1,..,n}$ and $X_{n+1,..,2n}$, and focus on the custodial symmetric cases (corresponding to a condition $|\lambda_-/\lambda_3|=2$ as in Ref.~\cite{Cai:2017wdu}), then $S$, $T$, and $U$ are not affected for the same scalar masses and the same potential parameters. On the other hand, denoting $g_D$ to be the new $\mathrm{U(1)}$ gauge coupling, a relatively large $\epsilon_W\sim 0.01\cdot ng_D(v^2/m_X^2)$ can be achieved when $n g_D$ is large.

After the SM Higgs field $H$ acquires the VEV $\hat{v}$ as usual,
\begin{eqnarray}
H = \begin{pmatrix}
i \phi^+ \\
\frac{\hat{v}+h+i \phi^0}{\sqrt{2}}
\end{pmatrix},
\end{eqnarray}
where $\hat{v}\approx246$ GeV, the kinetic mixing terms can be re-parametrized to be
\begin{eqnarray}
\mathcal{L}_{\rm eff} \supset  &-& \frac{ \epsilon_B}{2} \hat{Z}^{\prime}_{\mu \nu} B^{\mu \nu} -\frac{\epsilon_W}{2} \hat{Z}^{\prime}_{\mu \nu} W^{3 \mu \nu} - \frac{\epsilon_{BW}}{2} B_{\mu \nu} W^{3 \mu \nu} 
- \frac{\epsilon_{WW}}{4}  W^{3 \mu \nu} W^3_{\mu \nu}, \label{KineticMixingTerms}
\end{eqnarray}
where $\epsilon_W\equiv-\hat{v}^2/(2\Lambda_W^2),~\epsilon_{BW}\equiv-\hat{v}^2/(2\Lambda_{BW}^2)$, and $\epsilon_{WW}\equiv\hat{v}^4/(4\Lambda_{WW}^4)$.
Besides the kinetic mixing terms above, the vector bosons might also receive the mass corrections induced by
\begin{eqnarray}
\mathcal{L}_{\rm eff} \supset \delta m^2 (\hat{g}^{\prime}  \hat{Z}^{\prime}_{\mu} B^{\mu} - \hat{g} \hat{Z}^{\prime}_{\mu} W^{3 \mu}) + \frac{1}{4} (\hat{v}^2 + \delta v^2) (\hat{g}^{\prime 2} B_{\mu} B^{\mu} - 2 \hat{g} \hat{g}^{\prime} B_{\mu} W^{3 \mu} + \hat{g}^{\prime 2} W^3_{\mu} W^{3 \mu}).
\end{eqnarray}
where $\delta m^2\equiv-\lambda_{HZ'}\hat{v}^2/2$ and $\delta v^2\equiv \hat{v}^4/(4\Lambda_{HD}^2)$.
Combined with (\ref{LOriginal}), all the mass terms are given by
\begin{eqnarray}
\mathcal{L}_{mass}&=& \begin{pmatrix}
\hat{Z}^{\prime}_{\mu}, & B_{\mu}, &W^3_{\mu}
\end{pmatrix} \mathcal{M}_V^2 \begin{pmatrix}
\hat{Z}^{\prime \mu} \\
B^{\mu} \\
W^{3 \mu}
\end{pmatrix},
\\
\mathcal{M}_V^2&=& \begin{pmatrix}
\hat{m}_{Z^{\prime}}^2 & \hat{g}^{\prime} \delta m^2 & -\hat{g} \delta m^2 \\
\hat{g}^{\prime} \delta m^2  &  \frac{\hat{g}^{\prime 2}}{4} (\hat{v}^2 + \delta v^2)&  -\frac{\hat{g}^{\prime} \hat{g}}{4} (\hat{v}^2 + \delta v^2)\\
-\hat{g} \delta m^2 &  -\frac{\hat{g}^{\prime} \hat{g}}{4} (\hat{v}^2 + \delta v^2) & \frac{\hat{g}^2}{4} (\hat{v}^2 + \delta v^2)
\end{pmatrix}. \label{MassOrigin}
\end{eqnarray}
Before diagonalizing the mass-squared matrix (\ref{MassOrigin}), we have to diagonalize the kinetic terms
\begin{eqnarray}
\mathcal{L}_\mathrm{kin}&=&-\frac{1}{4} \begin{pmatrix}
\hat{Z}^{\prime}_{\mu \nu}, & B_{\mu \nu}, & W^3_{\mu \nu}
\end{pmatrix}\mathcal{K}_V \begin{pmatrix}
\hat{Z}^{\prime \mu \nu} \\
B^{\mu \nu} \\
W^{3 \mu \nu}
\end{pmatrix},
\\
\mathcal{K}_V&=& \begin{pmatrix}
1 & \epsilon_B & \epsilon_W \\
\epsilon_B  &  1 + \epsilon_{WW} &  \epsilon_{BW} \\
\epsilon_W &  \epsilon_{BW} & 1
\end{pmatrix}. \label{KOrigin}
\end{eqnarray}
To achieve this, we initially use a congruent transformation matrix composed of three elementary transformation matrices,
\begin{eqnarray}
V_C = V_1 V_2 V_3,\quad
V_C^\mathrm{T} \mathcal{K}_V V_C = I_{3 \times 3},
\end{eqnarray}
where
\begin{eqnarray}
V_1 &=& \begin{pmatrix}
1 & -\epsilon_B & -\epsilon_W \\
0 & 1 & 0 \\
0 & 0 & 1
\end{pmatrix}, ~V_2 =  \begin{pmatrix}
1 & 0 & 0 \\
0 & 1 & \frac{-\epsilon_{BW} + \epsilon_B \epsilon_W}{1-\epsilon_B^2}  \\
0 & 0 & 1
\end{pmatrix},
\nonumber \\
V_3 &=& \begin{pmatrix}
1 & 0 & 0 \\
0 & \frac{1}{\sqrt{1-\epsilon_B^2}} & 0 \\
0 & 0 & \sqrt{\frac{1-\epsilon_B^2}{1+\epsilon_{WW}-\epsilon_B^2-\epsilon_W^2-\epsilon_{BW}^2-\epsilon_B^2\epsilon_{WW}+2\epsilon_B\epsilon_W\epsilon_{BW}}}
\end{pmatrix}.
\end{eqnarray}
Correspondingly, the mass-squared matrix becomes
\begin{eqnarray}
(V_1 V_2 V_3)^\mathrm{T} \mathcal{M}_V^2 V_1 V_2 V_3. \label{Normalized}
\end{eqnarray}
Since
\begin{eqnarray}
\det[(V_1 V_2 V_3)^\mathrm{T} \mathcal{M}_V^2 V_1 V_2 V_3] = \det(\mathcal{M}_V^2) \det(V_1 V_2 V_3)^2 = 0,
\end{eqnarray}
we have one massless eigenstate identified to be exactly the physical photon. 

Then we are going to diagonalize the mass-squared matrix (\ref{Normalized}). Since the analytic solution is too hard for one to manipulate, though it does exist, we utilize a perturbative method to deal with it. We use the familiar EW rotation matrix
\begin{eqnarray}
V_{\text{SM}} = \begin{pmatrix}
1 & 0 & 0 \\
0 & -\frac{\hat{g}^{\prime}}{\sqrt{\hat{g}^{\prime 2} + \hat{g}^2}} & \frac{\hat{g}}{\sqrt{\hat{g}^{\prime 2} + \hat{g}^2}} \\
0 & \frac{\hat{g}}{\sqrt{\hat{g}^{\prime 2} + \hat{g}^2}} & \frac{\hat{g}^{\prime}}{\sqrt{\hat{g}^{\prime 2} + \hat{g}^2}}
\end{pmatrix}
\end{eqnarray}
to operate (\ref{Normalized}),
\begin{eqnarray}
(V_1 V_2 V_3 V_{\text{SM}})^\mathrm{T} \mathcal{M}_V^2 V_1 V_2 V_3 V_{\text{SM}} =\mathcal{M}_{0\text{d}}^2 + \delta\mathcal{M}^2,
\end{eqnarray}
where 
\begin{eqnarray}
\mathcal{M}_{0\text{d}}^2=\operatorname{diag}\left(\hat{m}_{Z^{\prime}}^2,~ (\hat{g}^2+\hat{g}^{\prime 2})\frac{\hat{v}^2+\delta v^2}{4},~ 0\right),
\end{eqnarray}
whose non-zero diagonal elements are much larger than the elements in $\delta\mathcal{M}^2$. Up to the second order of the perturbation theory, we have
\begin{equation}
V_{\text{f}, ii} \simeq 1-\sum_{i \neq j} \frac{\delta\mathcal{M}_{ij}^2\delta\mathcal{M}_{ji}^2}{2(\mathcal{M}_{0\text{d}, jj}^2 - \mathcal{M}_{0\text{d}, ii}^2)^2  },
\end{equation}
and
\begin{eqnarray}
V_{\text{f}, ij} &\simeq& \frac{\delta\mathcal{M}_{ji}^2}{\mathcal{M}_{0\text{d}, ii}^2-\mathcal{M}_{0\text{d}, jj}^2} + \sum_{k \neq i,j} \frac{\delta\mathcal{M}_{ik}^2 \delta\mathcal{M}_{ki}^2}{(\mathcal{M}_{0\text{d}, ii}^2-\mathcal{M}_{0\text{d}, jj}^2)(\mathcal{M}_{0\text{d}, ii}^2-\mathcal{M}_{0\text{d}, kk}^2)} \nonumber \\
&-& \frac{\delta\mathcal{M}_{ii}^2 \delta\mathcal{M}_{ji}^2}{(\mathcal{M}_{0\text{d}, ii}^2-\mathcal{M}_{0\text{d}, jj}^2)^2}
\end{eqnarray}
for $i \neq j$.
Finally, we acquire a transformation matrix
\begin{eqnarray}
V=V_1 V_2 V_3 V_{\text{SM}} V_{\text{f}}
\end{eqnarray}
to diagonalize $\mathcal{M}_V^2$,
\begin{eqnarray}
V^\mathrm{T} \mathcal{M}_V^2 V = \text{diag}(m_{Z^{\prime}}^{2},~ m_Z^2,~ 0),
\end{eqnarray}
where
\begin{eqnarray}
m_{Z^{\prime}}^{2} &=& \hat{m}_{Z^{\prime}}^{2} + O(\epsilon_{W,B,BW, WW}, \delta v^2, \delta m^2), \nonumber \\
m_{Z}^{2} &=& (\hat{g}^2+\hat{g}^{\prime 2})\frac{\hat{v}^2+\delta v^2}{4} + O(\epsilon_{W,B,BW, WW}, \delta v^2, \delta m^2) \label{PhysicalMassZ}
\end{eqnarray}
are physical masses squared for the mass eigenstates $Z^{\prime}$ and $Z$.

\section{Evaluations of oblique $S^\prime$, $T^\prime$, and $U^\prime$ parameters} \label{ObliqueEvaluation}

Usually, one regards the Fermi constant $G_F$, QCD coupling constant $\alpha_s(m_Z)$ defined at the $Z$ boson mass scale, the fine structure constant $\alpha(m_Z)$, the $Z$ boson pole mass $m_Z$, the top quark pole mass $m_t$, and the Higgs boson pole mass $m_h$ as the basic input parameters to the electroweak theories. The $W$ boson mass $m_W^{\mathrm{SM}}$ and the $Z$ boson decay parameters $R_{e, \mu,\tau}^{\mathrm{SM}}$, $A_{e, \mu, \tau}^{\mathrm{SM}}$, and $\Gamma_{Z}^{\mathrm{SM}}$ are then predicted for comparing with the experimental results. The deviation between the theoretical predictions and experimental results due to EW oblique corrections are usually summarized to be the three oblique parameters $S$, $T$, and $U$ (see the definitions in Eq.~\eqref{STUOriginal})~\cite{Peskin:1990zt, Peskin:1991sw}. However, the original $S$, $T$, $U$ only include the zeroth and first $p^2$-order corrections of the vacuum polarizations, which cannot fully describe the deviation induced by the mixings of $Z^\prime$. From the EFT point of view, we can integrate out the $Z^\prime$ boson at scales much lower than its mass and generate some dim-6, dim-8, and even dim-10 SMEFT operators which significantly distort the vacuum polarizations of EW gauge fields. We leave more detailed discussion of EFT in App.~\ref{Appendix}.


Before evaluating the effective $S^\prime$, $T^\prime$, and $U^\prime$, we need to clarify the definition of three ``physical" quantities in this work. The ``physical'' value of the Weinberg angle $\theta_w$ is defined by~\cite{Burgess:1993vc, Babu:1997st}
\begin{eqnarray}
s_w^2 c_w^2 = \frac{ \pi \alpha}{\sqrt{2} m_Z^2 G_F},
\end{eqnarray}
where $s_w \equiv \sin\theta_w$ and $c_w \equiv \cos\theta_w$.
The physical value of $m_Z$ has been defined in Eq.~(\ref{PhysicalMassZ}). The fine structure constant $\alpha$ is extracted from the effective coupling constant between the massless vector boson (photon) and the charged particles,
\begin{eqnarray}
\alpha &=& \frac{e^2}{4 \pi}, \nonumber \\
e &=& \frac{\hat{g}^{\prime}}{2} V_{23} + \frac{\hat{g}}{2} V_{33}.
\end{eqnarray}
The Fermi constant is defined as
\begin{eqnarray}
G_F = \frac{1}{\sqrt{2} v^2},
\end{eqnarray}
which is the only parameter that receives no new physics contribution in this paper.

Usually the effective $S^\prime$, $T^\prime$, and $U^\prime$ parameters can be extracted by directly calculating the self-energy diagrams of the EW gauge bosons. In this paper, we use another equivalent method. The neutral current (NC) and the charged current (CC) parameters extracted from the experimental results can be adopted for comparing with the theoretical predictions to work out the oblique parameter values. Following the steps in Ref.~\cite{Burgess:1993vc}, one can acquire $S^\prime$, $T^\prime$, and $U^\prime$ from the NC and CC coefficients. Expressed by the mixing parameters, the results of the $S^\prime$ and $T^\prime$ parameters are given by
\begin{eqnarray}
\alpha S^\prime &=& 4\left[ -\frac{V_{22}}{V_{23}} \frac{s_w c_w}{1+{\alpha T^\prime}/{2}} - s_w^2 (c_w^2-s_w^2) + s_w^2 c_w^2 \alpha T^\prime \right], \label{SParameter}\\
\alpha T^\prime &=& 2 s_w c_w \left(\frac{V_{32}}{V_{33}} - \frac{V_{22}}{V_{23}}\right) - 2. \label{TParameter}
\end{eqnarray}

The $U^\prime$ parameter should be extracted from the charged current coupling constants. However, equivalently it is more convenient to look into the $W$ boson mass~\cite{Peskin:1991sw}
\begin{eqnarray}
m_W = m_W^{\mathrm{SM}} \left[ 1-\frac{\alpha}{4 (c_w^2-s_w^2)} \left(S^\prime-2c_w^2T^\prime-\frac{c_w^2-s_w^2}{2s_w^2}U^\prime\right) \right], \label{WMass2U}
\end{eqnarray}
where $m_W = {\hat{g} v}/{2}$ is the physical $W$ boson mass. Neglecting the loop corrections, the SM prediction is $m_W^{\mathrm{SM}} = \frac{\sqrt{4 \pi \alpha}}{2 s_w \sqrt{\sqrt{2} G_F}}$. With the difference between $m_W$ and $m_W^{\mathrm{SM}}$, and the $S^\prime$, $T^\prime$ parameters acquired in Eqs.~(\ref{SParameter}) and (\ref{TParameter}), one can easily derive the $U^\prime$ parameter.

Here we list the expressions of $S^\prime$, $T^\prime$, and $\delta m_W^2 = m_W^2 - (m_W^{\mathrm{SM}})^2$ expanded up to the second order of the parameter set $\epsilon_{B, W, BW, WW}$, $\delta m^2$, and $\delta v^2$. The results are given by 
\begin{eqnarray}
\alpha S^\prime &=& \frac{4g g^{\prime}}{g^2 + g^{\prime 2}} \epsilon_{BW} - \frac{g^2 g^{\prime 2} v^2 (4 m_{Z^{\prime}}^2 - g^2 v^2)}{4 (g^2 + g^{\prime 2})(m_{Z^{\prime}}^2 - m_Z^2)^2} \epsilon_B^2 + \frac{g g^{\prime} v^2 [4 (g^2 + g^{\prime 2}) m_{Z^{\prime}}^2 - (g^4 + g^{\prime 4}) v^2]}{4 (g^2 + g^{\prime 2})(m_{Z^{\prime}}^2 - m_Z^2)^2} \epsilon_B \epsilon_W \nonumber \\
&-& \frac{g^2 g^{\prime 2} v^2 (4 m_{Z^{\prime}}^2 - g^{\prime 2} v^2)}{4 (g^2 + g^{\prime 2})(m_{Z^{\prime}}^2 - m_Z^2)^2} \epsilon_W^2  + \frac{ g^2 g^{\prime} [4 m_{Z^{\prime}}^2 - (g^2 - g^{\prime 2}) v^2]}{(g^2 + g^{\prime 2})( m_{Z^{\prime}}^2-m_Z^2)^2}  \epsilon_B \delta m^2 \nonumber \\
&-& \frac{ g g^{\prime 2} [4 m_{Z^{\prime}}^2 + (g^2 - g^{\prime 2}) v^2]}{(g^2 + g^{\prime 2})( m_{Z^{\prime}}^2-m_Z^2)^2} \epsilon_W \delta m^2 + \frac{4g^2 g^{\prime 2} (6 g^2 g^{\prime 2}-g^4-g^{\prime 4})}{(g^4 - g^{\prime 4})^2} \epsilon_{BW}^2 \nonumber \\
&-& \frac{4g g^{\prime 3}}{(g^2 + g^{\prime 2})^2} \epsilon_{WW}\epsilon_{BW}+\frac{8g^3g^{\prime 3}}{(g^2-g^{\prime 2})^2(g^2+g^{\prime 2})v^2}\epsilon_{BW}\delta v^2 + \frac{3 g^6 g^{\prime 2}- 2 g^4 g^{\prime 4} +  3 g^2 g^{\prime 6}}{ (g^4 - g^{\prime 4})^2 v^4} (\delta v^2)^2 \nonumber \\
&-& \frac{4 g^2 g^{\prime 2}}{ (g^2 + g^{\prime 2})(m_{Z^{\prime}}^2 - m_Z^2)^2} (\delta m^2)^2, \label{S_Expansion}
\end{eqnarray}
\begin{eqnarray}
\alpha T^\prime &=& -\frac{\delta v^2}{v^2} - \frac{ m_{Z^{\prime}}^2 v^2}{4 (m_Z^{\prime 2}-m_Z^2)^2} (g^{\prime} \epsilon_B - g \epsilon_W)^2 + \frac{2 m_Z^2}{ (m_{Z^{\prime}}^2 - m_Z^2)^2} ( g^{\prime} \epsilon_B  - g \epsilon_W )\delta m^2 \nonumber \\
&+& \frac{3}{4 v^4} (\delta v^2)^2 + \frac{4 ( m_{Z^{\prime}}^2 - 2 m_Z^2)}{ (m_{Z^{\prime}}^2 - m_Z^2)^2 v^2} (\delta m^2)^2, \label{T_Expansion}
\end{eqnarray}
\begin{eqnarray}
\delta m_W^2 &=& -\frac{g^3 g^{\prime} v^2}{2 (g^2 -  g^{\prime 2})} \epsilon_{BW} - \frac{g^4}{4 (g^2 -  g^{\prime 2}) } \delta v^2 +\frac{g^2 v^2}{4} \epsilon_{WW}
\nonumber \\
&+& \frac{g^4 v^4}{16 (g^2-g^{\prime 2})(m_{Z^{\prime}}^2 - m_Z^2)} (g^{\prime} \epsilon_B - g \epsilon_W)^2
+ \frac{2 g^4 v^2}{4 (g^2 - g^{\prime 2}) (m_{Z^{\prime}}^2 - m_Z^2)} (g \epsilon_W - g^{\prime} \epsilon_B)\delta m^2
\nonumber \\
&-& \frac{g^4 (g^2 - 3 g^{\prime 2})(g^2 + g^{\prime 2}) v^2}{4 (g^2 - g^{\prime 2})^3} \epsilon_{BW}^2
+ \frac{g^3 g^{\prime} (g^2 - 2 g^{\prime 2}) v^2}{2 (g^2 - g^{\prime 2})^2} \epsilon_{BW} \epsilon_{WW}\nonumber \\
&-& \frac{g^2v^2 }{4} \epsilon_{WW}^2 +\frac{g^4(g^2-2g^{\prime 2})}{4(g^2-g^{\prime 2})^2}\epsilon_{WW}\delta v^2+ \frac{g^4 g^{\prime 4}}{4 (g^2 - g^{\prime 2})^3 v^2} (\delta v^2)^2 \nonumber \\
&-&\frac{g^5g^\prime (g^2-3g^{\prime 2})}{2(g^2-g^{\prime 2})^3}\epsilon_{BW}\delta v^2 +\frac{ g^4}{ (g^2 - g^{\prime 2})(m_{Z^{\prime}}^2 - m_Z^2)} (\delta m^2)^2. \label{DeltaW_Expansion}
\end{eqnarray}

Originally, the above $g$, $g^{\prime}$, $v$, etc. parameters should be ``hatted'' and become $\hat{g}$, $\hat{g}^{\prime}$, $\hat{v}$, etc.  However, since the shifts of all these parameters from the physical ones are extremely small, we can conveniently utilize the physical parameters to evaluate the $S^\prime$, $T^\prime$, and $\delta m_W^2$ instead.

Since the above formulas for evaluating the oblique parameters are rather complicated, it is not essential to perform a thorough fitting on all these parameters, so in the rest of this paper we will focus on several specific cases.

\section{Phenomenological discussions on $Z^{\prime}$ collider bounds and the oblique parameters} \label{ColliderBounds}

\subsection{$Z^{\prime}$ collider bounds}

In order to generate a significant positive $U^{\prime}$ parameter, the $Z^\prime$ mass should lie within a range $m_Z < m_{Z^\prime}\lesssim 400$~GeV. Since $Z^\prime$ couples to both leptons and quarks due to its kinetic mixing with SM gauge fields, it can be produced both in lepton and hadron colliders. The neutral current interactions with $Z'$ are
\begin{eqnarray}
    \mathcal{L}_{Z'_\mu J_{f}^\mu}=\sum_f\bar{f}\gamma^\mu\left[g_{f}^{(V)}+  g_{f}^{(A)}\gamma_5 \right]fZ^\prime_\mu,
\end{eqnarray}
where $f=u_i,d_i,\nu_i,e_i$ are SM fermions, and the couplings are given by
\begin{eqnarray}
g_{f}^{(V)}&\approx&e\left[Q_f\frac{s_w\epsilon_W+(c_w-r/c_w)\epsilon_B+t_w\xi}{r-1}-\frac{r(c_w\epsilon_W-s_w\epsilon_B)+\xi}{2s_w c_w(r-1)}T_{f_L}^3\right],\\
g_{f}^{(A)}&\approx&Q_fe\frac{r(c_w\epsilon_W-s_w\epsilon_B)+\xi}{2s_w c_w(r-1)}T_{f_L}^3,
\end{eqnarray}
with $t_w \equiv \tan\theta_w$, $r\equiv m_{Z'}^2/m_Z^2$, and $\xi\equiv  \sqrt{g^2+g'^2}\delta m^2/m_{Z}^2$.

For $m_{Z'}\lesssim 209$~GeV, $Z^{\prime}$ may be directly produced at the LEP collider with a significant signal. The null result of the on-shell $Z^{\prime}$ searches at the LEP either pushes the $Z'$ mass heavier than $209$~GeV, or suppresses the couplings $g_l^{(V,A)}$ to leptons smaller than $\mathcal{O}(10^{-2})$~\cite{ParticleDataGroup:2020ssz}. For $m_{Z'}>209$~GeV, the LEP bound on the off-shell $Z^{\prime}$ production can be interpreted to be $m_{Z'}/\sqrt{g_e^{(L)}g_f^{(L)}}\gtrsim4$~TeV~\cite{Carena:2004xs}, where $g_f^{(L)}=(g_f^{(V)}-g_f^{(A)})/2$. In order to avoid the stringent on-shell bound from the LEP, we only concern in the off-shell range $m_{Z'}>220$ GeV, so  $g_f^{(L)}$ is calculated to be $\lesssim 0.03$  within our interested parameter region $|\epsilon_{B,W}|<0.1$ when $\xi$ is negligible. Therefore we do not have to worry about the LEP bounds in this paper.

At the hadron colliders, $Z'$ might be probed through the $pp\to Z'\to \text{dijets}$ searches. Current bounds on the universal vector-current coupling of $Z'$ to quarks is $g'_q\lesssim0.1$ for $220~\textrm{GeV}\lesssim m_{Z'}\lesssim 400$~GeV~\cite{CMS:2019emo,UA2:1993tee,CMS:2018kcg}. A more stringent estimated bound $g'_q\lesssim0.05$ can be acquired around $m_{Z'}\sim$TeV~\cite{CMS:2016ltu,ATLAS:2019fgd} (see Figure 88.2 in Ref.~\cite{ParticleDataGroup:2020ssz} for a summary of hadron collider bounds). Since our $Z'$-fermion couplings are chiral, in order to compare with the $g'_q$ bounds, we define an effective coupling
\begin{eqnarray}
    g_{q}^\text{eff}=\sqrt{\frac{(g_q^{(V)})^2 + (g_q^{(A)})^2 }{2}} 
\end{eqnarray}
as an estimate of our theoretical value for $g'_q$. We find that $g_{q}^\text{eff}\lesssim 0.04$ for $|\epsilon_{B,W}|<0.1$ and $m_{Z'}>209$~GeV with negligible $\xi$ is still consistent with the current LHC bounds. Since the parameter regions of our interest are sufficiently safe from the analysis of the collider constraints, we shall neglect them in our following discussions.

\subsection{Bounds on the oblique parameters $W$ and $Y$}

As will be discussed in the Appendix~\ref{Appendix}, the contributions from the $\epsilon_{W, B}$ terms to $S^{\prime}$, $T^{\prime}$, and $U^{\prime}$ include the contributions from $V$, $W$, $X$, and $Y$. (See their definitions in Appendix~\ref{Appendix}.) To constrain these parameters, one has to study the low-energy experimental data as in Refs.~\cite{Barbieri:2004qk,Cacciapaglia:2006pk}, or directly extract the shape of the vector boson propagators through the collider data. In this paper we only discuss the latter constraint, which is more stringent. Refs.~\cite{Torre:2020aiz,Panico:2021vav} provided the proposal to utilize the charged and neutral Drell-Yan differential cross-section measurements to constrain the $W$ and $Y$ parameters. In Ref.~\cite{CMS:2022yjm}, the CMS collaboration published a measurement result $W=-1.2^{+0.5}_{-0.6}\times10^{-4}$ through charged Drell-Yan processes. Ref.~\cite{Strumia:2022qkt} applied the $l^{+} l^{-}$ data presented in Ref.~\cite{ATLAS:2017fih} to constrain the $Y$ parameter as $|Y|\lesssim 2\times10^{-4}$. Such stringent bounds on $W$ and $Y$ naively exclude nearly the whole interested parameter space. However, the fitting results are actually based upon some assumptions which are not the case in this paper.

In the above studies, the $W$ parameter is extracted from the charged Drell-Yan data, and the corresponding effective operator is $(D_{\rho} W_{\mu \nu}^a)^2$, which affect $W^3$ and $W^{\pm}$ universally. This is eligible when $V=0$. In our paper, however, the kinetic mixing term $\frac{\epsilon_W}{2} \hat{Z}^{\prime}_{\mu \nu} W^{3 \mu \nu}$ in the Lagrangian \eqref{KineticMixingTerms} as well as its corresponding effective operator $\frac{1}{2 \Lambda_{W}^2 } \hat{Z}^{\prime}_{\mu \nu} W^{a \mu \nu} H^{\dagger} \sigma^a H$ in the Lagrangian  \eqref{LOriginal} only affects $W^3$ without disturbing $W^{\pm}$, resulting in $W=V$. (We follow Ref.\cite{Barbieri:2004qk} to define $W$ by $\Pi''_{W^3W^3}(0)$, and define $V$ by $\Pi''_{W^3W^3}(0)-\Pi''_{W^+W^-}(0)$.) Since in this case $W$ does not correct the charged Drell-Yan processes mediated by the off-shell $W^{\pm}$, the $W$ bound presented in Ref.~\cite{CMS:2022yjm} can be safely neglected.

Fig.~8 in Ref.~\cite{Torre:2020aiz} showed the projected exclusion regions of the $W$ and $Y$ parameters from both the neutral and charged Drell-Yan measurements at 13~TeV LHC. As we have mentioned, the charged result does not constrain our case, so only the neutral results are effective. The combined constraints on $W$ and $Y$ in Ref.~\cite{Strumia:2022qkt} are also based upon the $V=0$ assumption, and therefore become invalid again.

In the $Y=0$ case at the 13~TeV LHC with an integrated luminosity $100~\si{fb^{-1}}$, the neutral results in Ref.~\cite{Torre:2020aiz} predicted a $95\%$ CL bound $|W| \lesssim 0.4 \times 10^{-3}$, which is equivalent to $|U'| \lesssim 0.1$. One can verify from the figures in the next section that such a projected bound is at the brink of our desired parameter space to accommodate an appropriate $\delta m_W^2$ when $\epsilon_B = 0$. However, this is only a theoretical estimation, and up till now, we find no extractions of the $W$, $Y$ constraints merely from the neutral Drell-Yan experimental data in the literature. Moreover, moderate $\epsilon_{B}$ and $\epsilon_{W}$ should also give rise to a non-negligible $X$, which modifies the neutral boson propagators as well, but is simply discarded in all the references above. Therefore all the existing collider bounds on the $W$ and $Y$ parameters become inapplicable in our case, except for a small region near $\epsilon_B =0$. Thus, we neglect all of them in our following discussions.

\section{Numerical results} \label{NumericalResults}

In order to study the space of the  parameters, we adopt the ``standard average'' result of the $S$, $T$, $U$ parameters from the EW global fit with the recent CDF $m_W$ measurement in Ref.~\cite{deBlas:2022hdk}, as tabulated in Tab.~\ref{STU_Fitted}. In Ref.~\cite{deBlas:2022hdk} only the measurements of the precision observables at the EW scale and the Fermi constant $G_F$ are included, permitting a straightforward comparison with our effective $S^{\prime}$, $T^{\prime}$ and $U^{\prime}$.  For each of the parameter point, we compute the oblique parameters and then evaluate the corresponding $\chi^2$ based on this result.

\begin{table}[!t]
\setlength{\tabcolsep}{.5em}
\begin{tabular}{|c|c|ccc|}
\hline
  &  Result & \multicolumn{3}{c|}{Correlation} \\
\hline
$S$ & $0.005 \pm 0.097$ & $1.00$ & & \\
$T$ & $0.04 \pm 0.12$ & $0.91$ & $1.00$ & \\
$U$ & $0.134 \pm 0.087$ & $-0.65$ & $-0.88$ & $1.00$ \\
\hline
\end{tabular}
\caption{Global fit results of the oblique parameters $S$, $T$, and $U$ adopted from Ref.~\cite{deBlas:2022hdk}.} \label{STU_Fitted}
\end{table}

\begin{figure}[!t]
\begin{tabular}{ll}
    \includegraphics[width=0.43\textwidth]{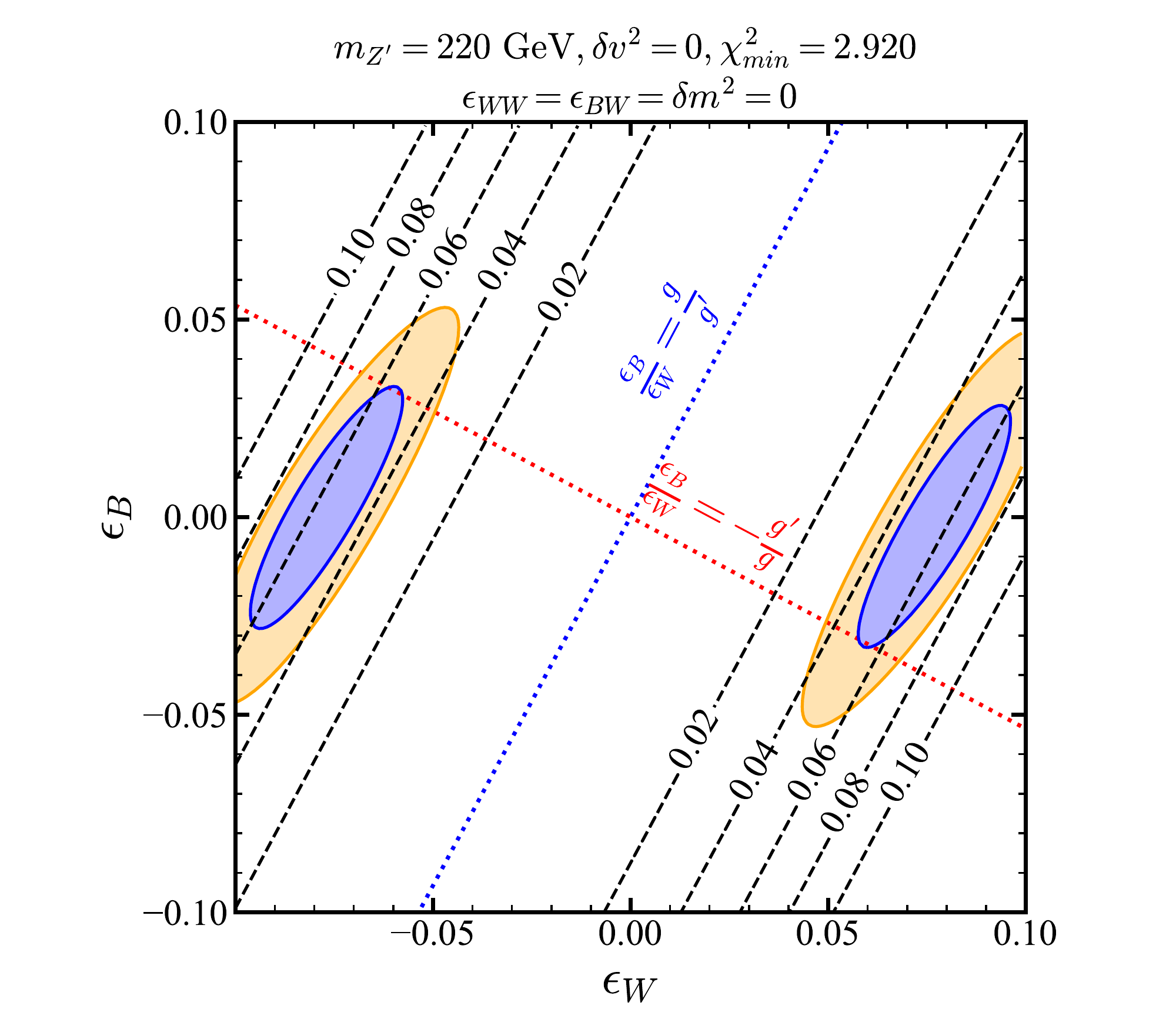}
    & \includegraphics[width=0.48\textwidth]{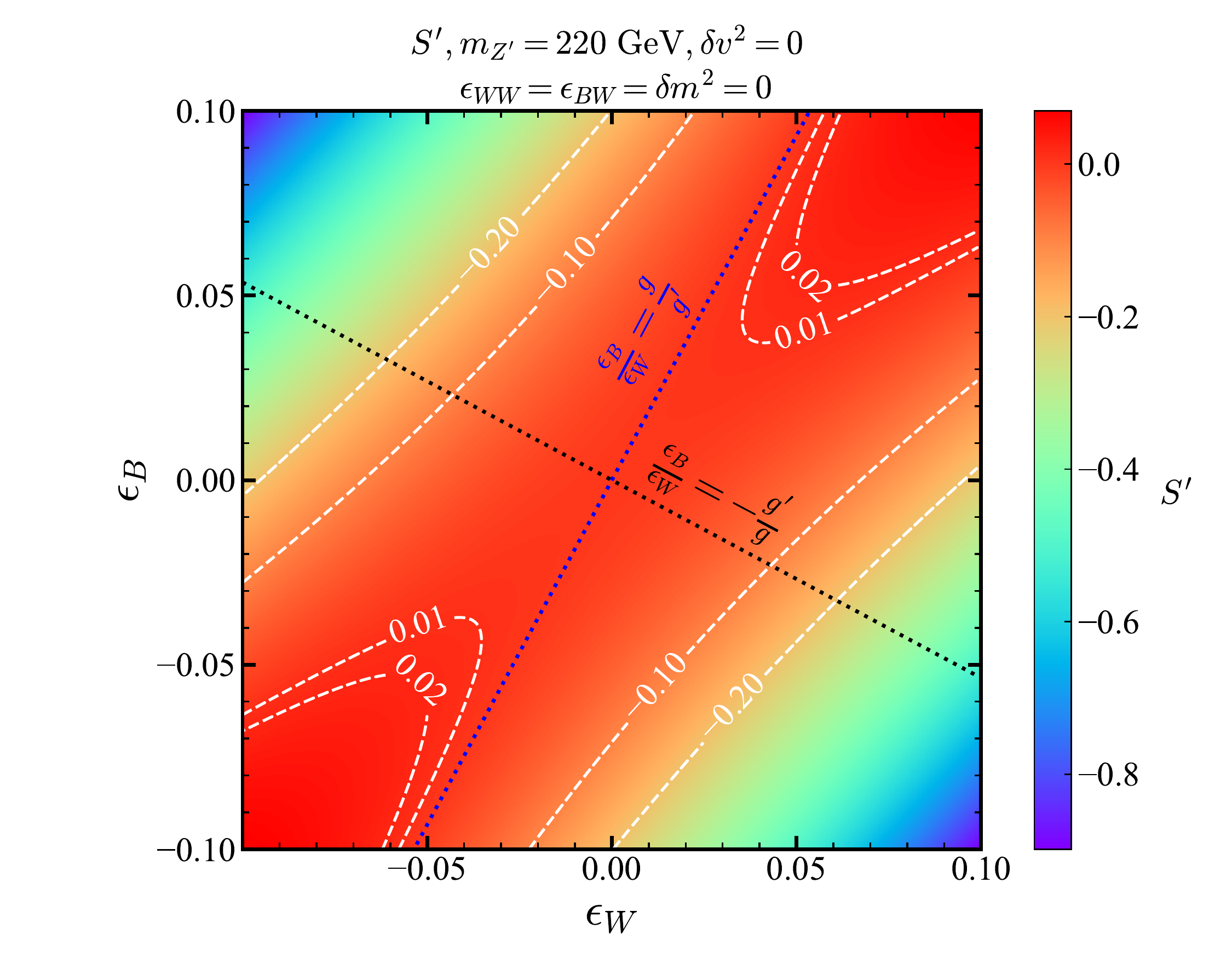} \\
    \includegraphics[width=0.48\textwidth]{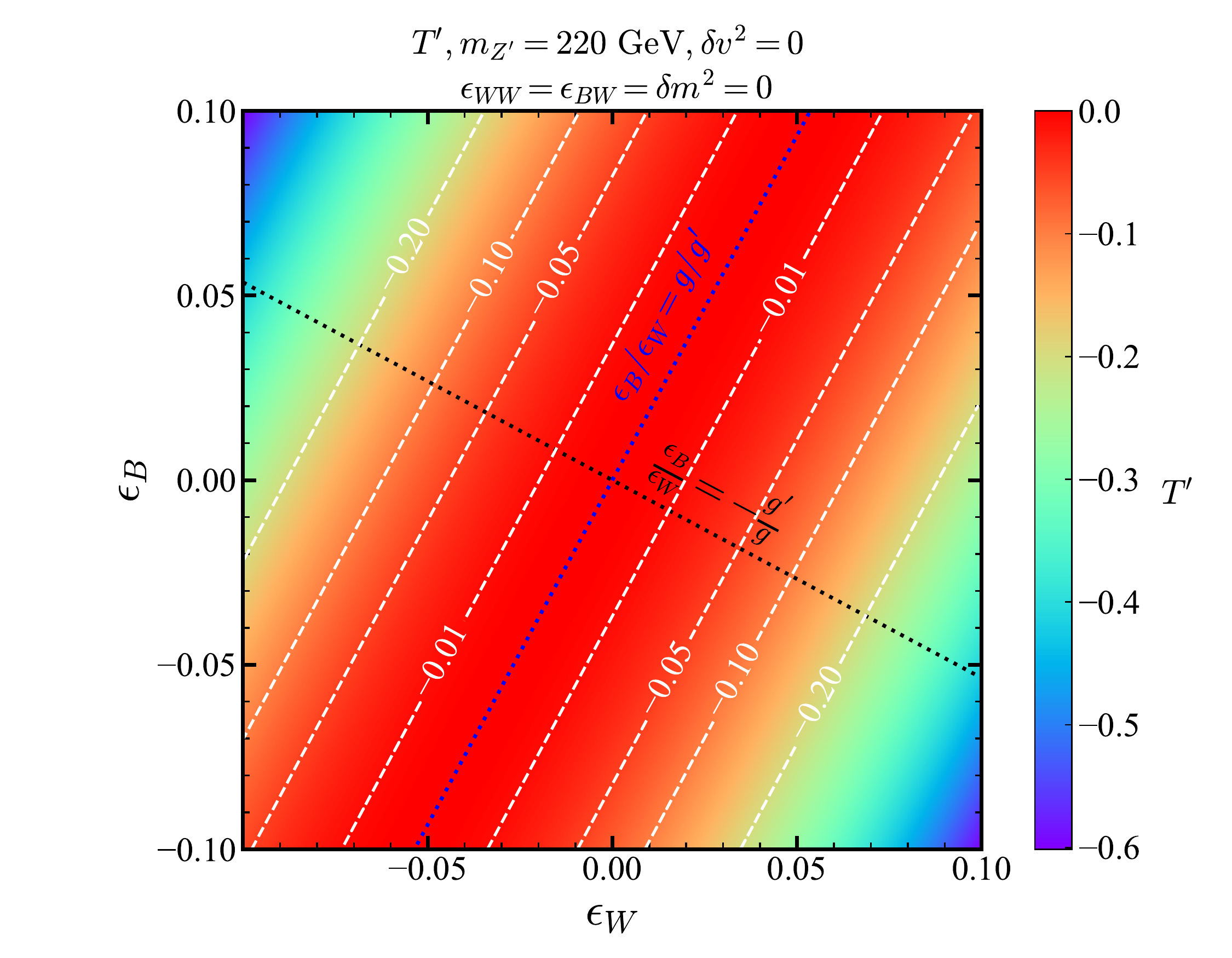}
    & \includegraphics[width=0.48\textwidth]{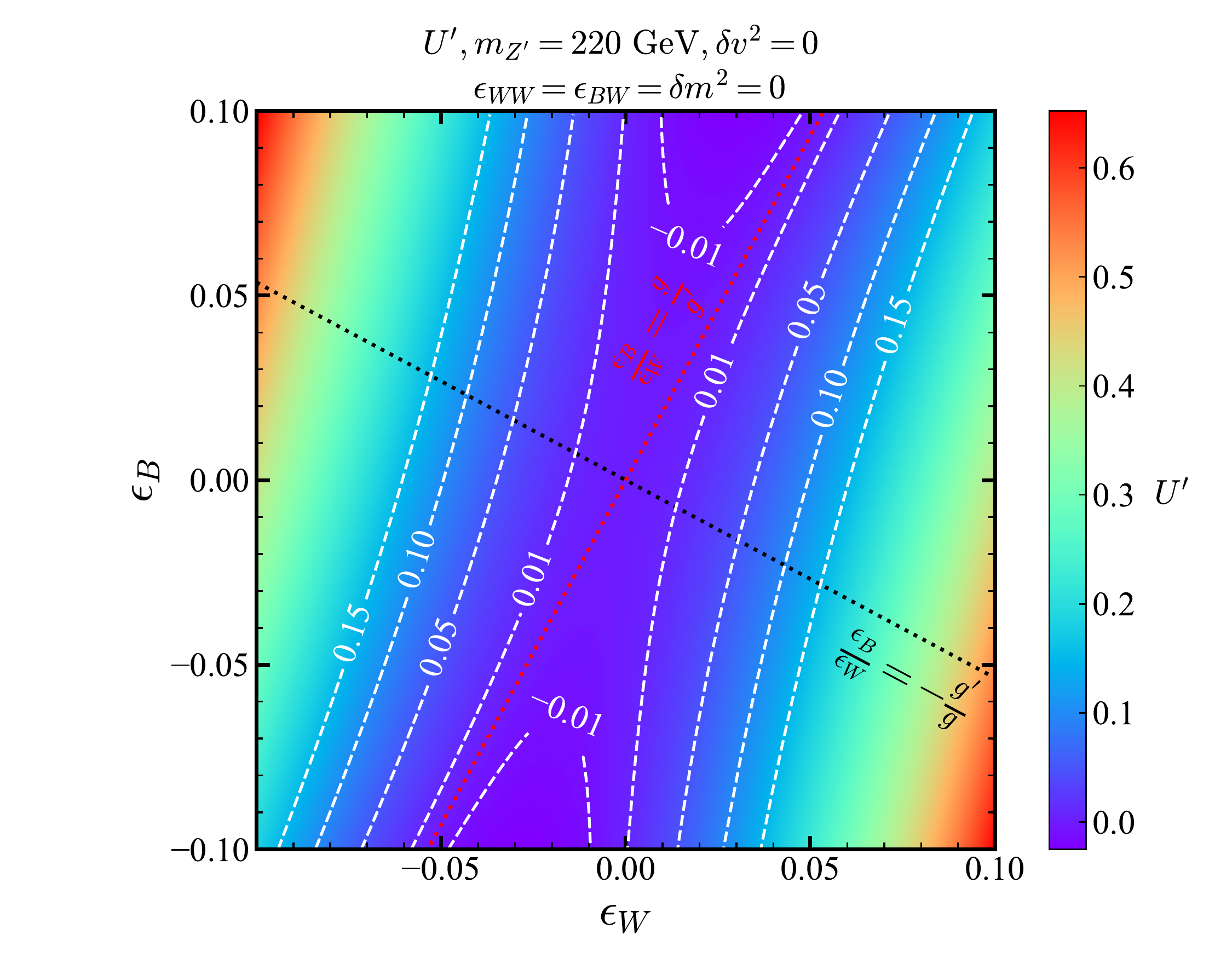}
\end{tabular}
    \caption{Plots for $m_{Z^{\prime}}=220$~GeV and $\delta v^2  = \epsilon_{WW} = \epsilon_{BW} = \delta m^2 = 0$ on the $\epsilon_B$ versus $\epsilon_W$ plain. The upper left panel shows the $1\sigma$ (blue), $2\sigma$ (orange) regions calculated according to the global fit result in Ref.~\cite{deBlas:2022hdk}, as well as the contours of $\delta m_W = m_W - m_W^\mathrm{SM}$ also displayed. The 
   marks ``0.02'', ``0.04'', etc. correspond to $\delta m_W = 0.02, 0.04$ GeV, etc. The minimal chi-squared $\chi^2_{\min}$ is indicated in the plot title. The remaining three panels display the contours of $S^{\prime}$, $T^{\prime}$, and $U^{\prime}$. In all the panels, the ${\epsilon_B}/{\epsilon_W} = {g}/{g^{\prime}}$ and ${\epsilon_B}/{\epsilon_W} =-{g}/{g^{\prime}}$ lines indicate the photon-$Z^{\prime}$ and $Z$-$Z^{\prime}$ mixings, respectively.}
    \label{mZ_220_NoDeltaV2}
\end{figure}

\begin{figure}[!t]
\begin{tabular}{ll}
    \includegraphics[width=0.43\textwidth]{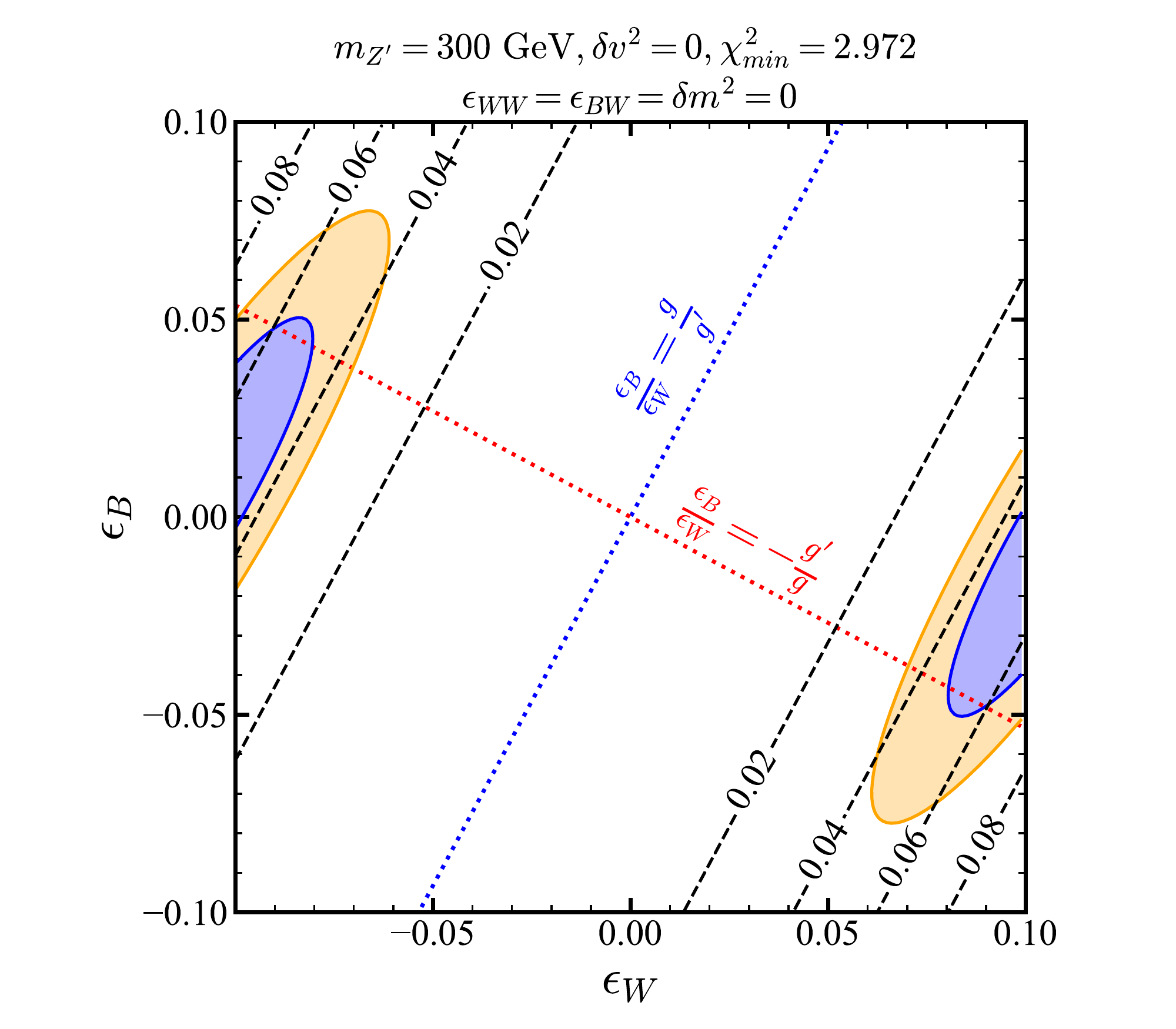} &
    \includegraphics[width=0.48\textwidth]{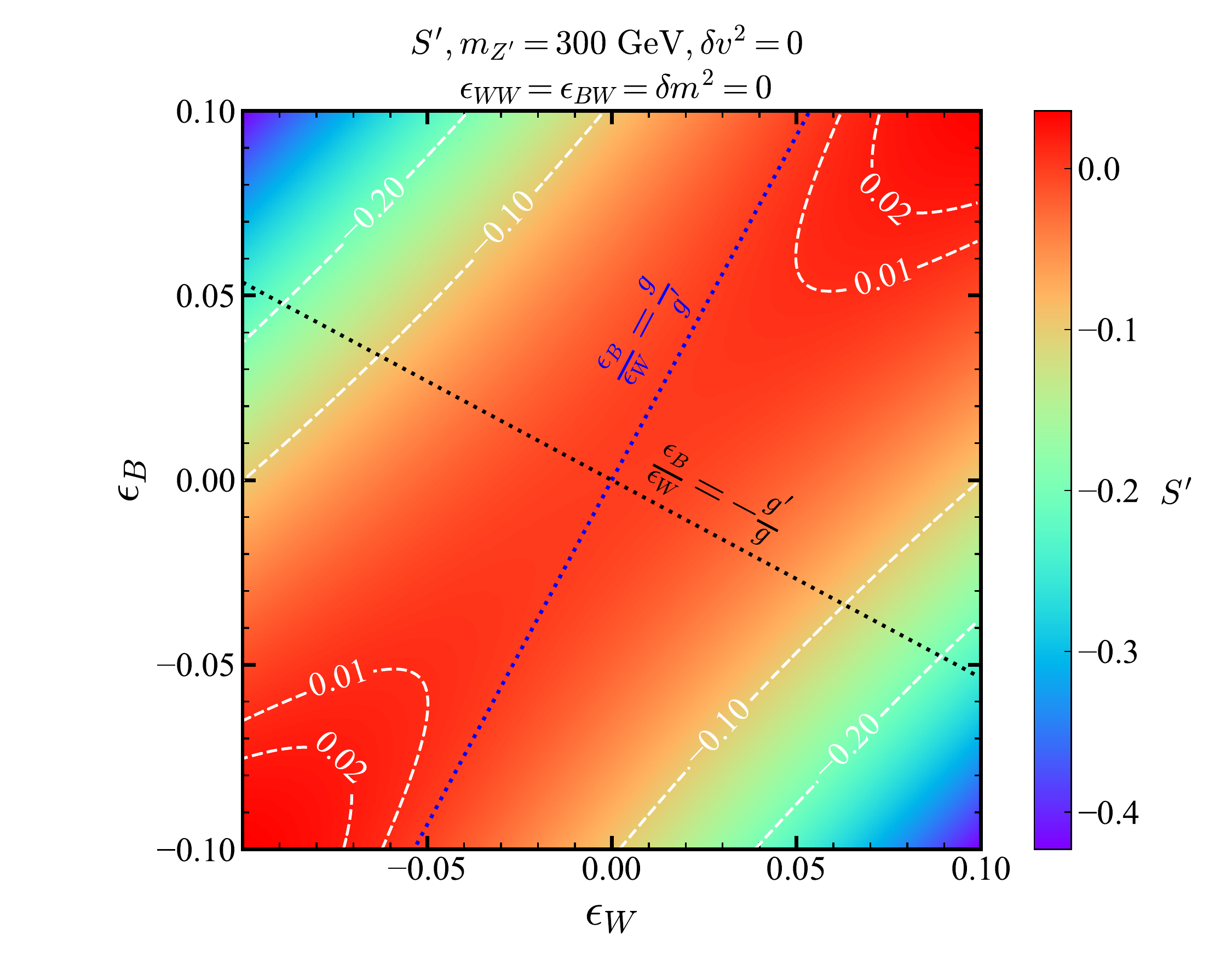} \\
    \includegraphics[width=0.48\textwidth]{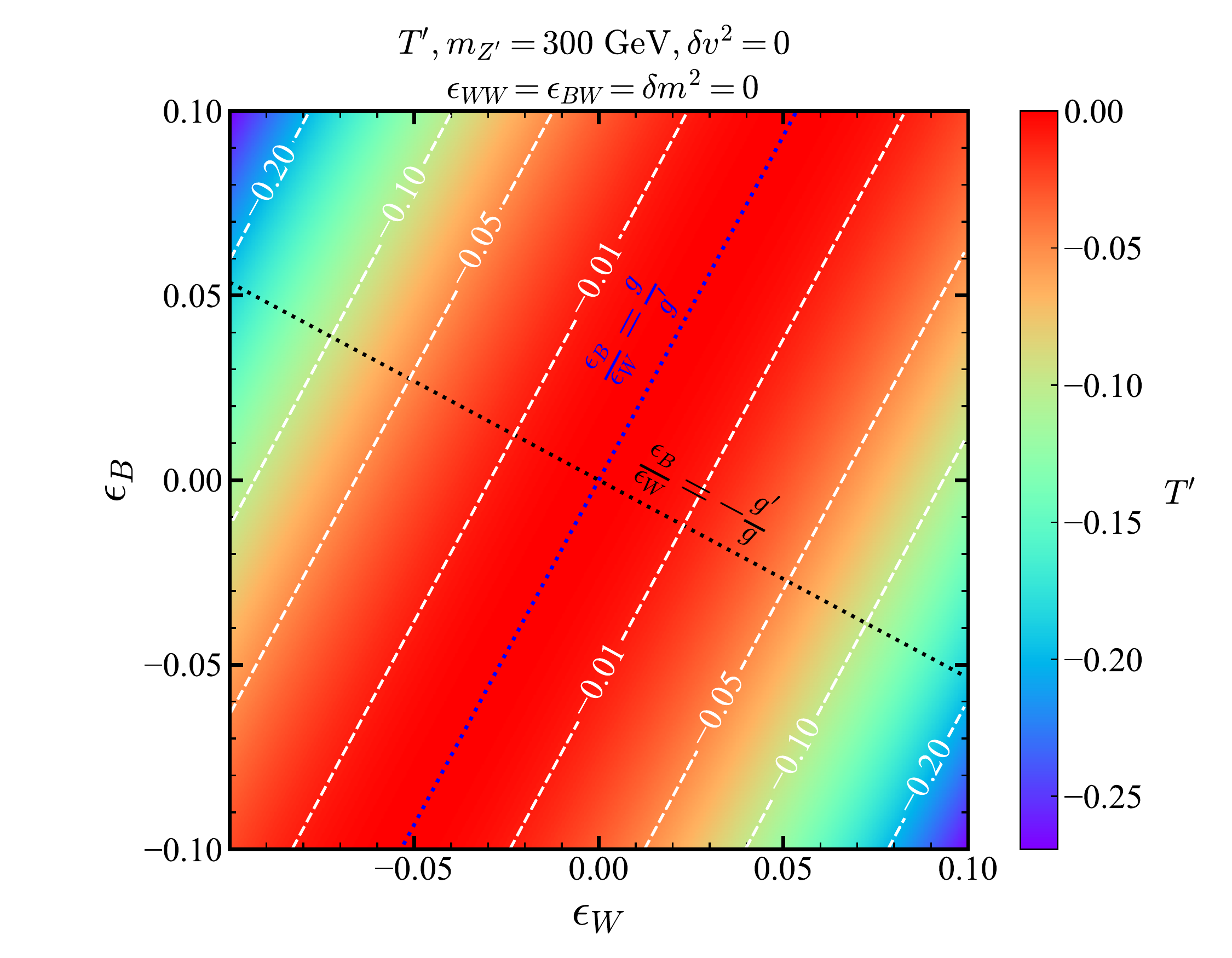} &
    \includegraphics[width=0.48\textwidth]{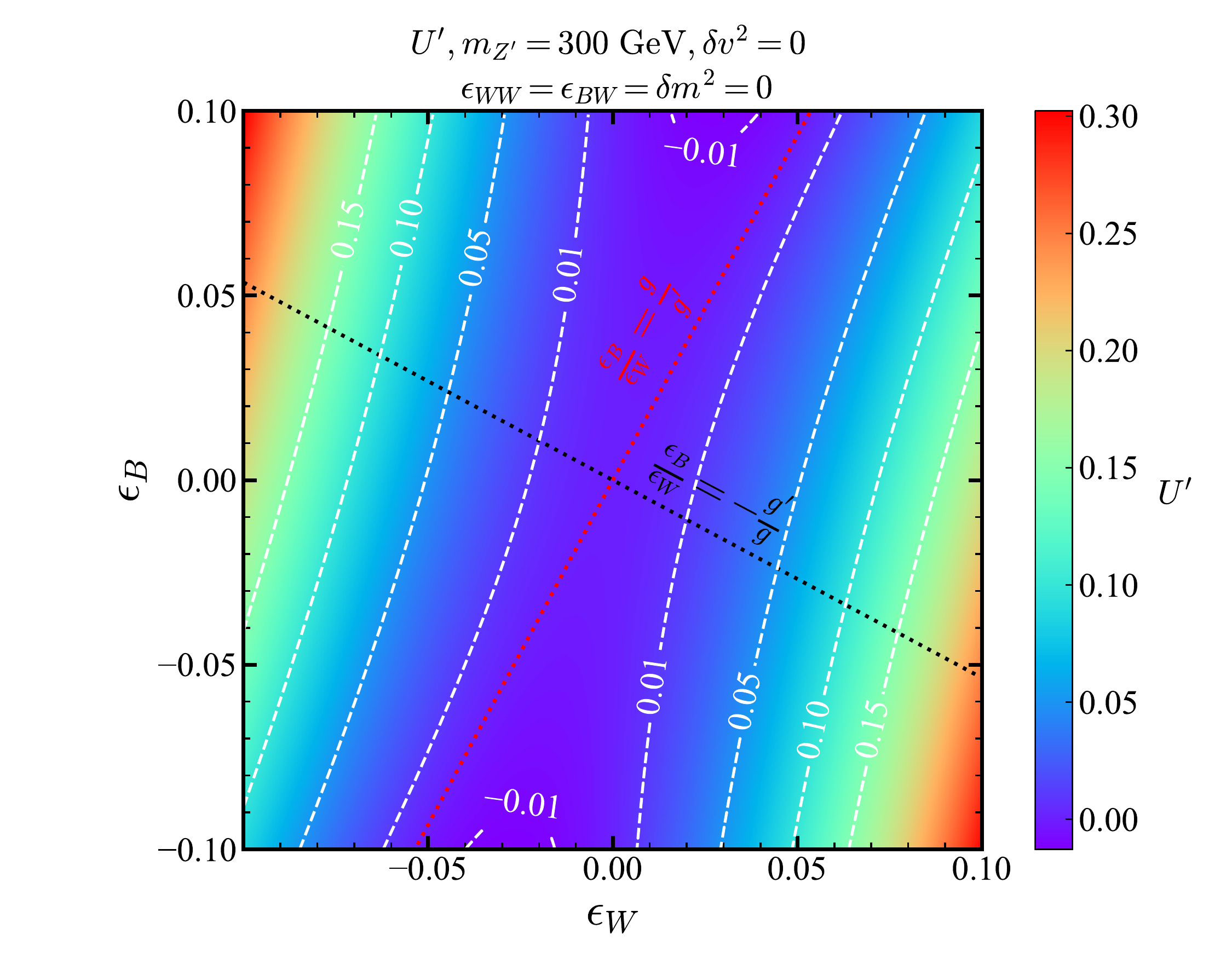}
\end{tabular}
    \caption{Plots for $m_{Z^{\prime}}=300$~GeV and $\delta v^2  = \epsilon_{WW} = \epsilon_{BW} = \delta m^2 = 0$ on the $\epsilon_B$ versus $\epsilon_W$ plain. The symbols are the same as in Fig.~\ref{mZ_220_NoDeltaV2}.}
    \label{mZ_300_NoDeltaV2}
\end{figure}

The kinetic mixing parameters $\epsilon_{B}$ and $\epsilon_{W}$ contribute positive values to the $U^{\prime}$ parameter, lifting the mass of the $W$ boson. However, from Eq.~(\ref{T_Expansion}) we learn that the $T^{\prime}$ parameter simultaneously acquires a negative contribution. Therefore, a tension arises when we try to fit with the results in Tab.~\ref{STU_Fitted} in the case that all the other mixing parameters disappear. In Figs.~\ref{mZ_220_NoDeltaV2} and \ref{mZ_300_NoDeltaV2}, we plot the fit results on the $\epsilon_B$ versus $\epsilon_W$ plain for $m_{Z^{\prime}}=220$~GeV and 300~GeV, respectively. The values of the oblique parameters are also shown as contours. The non-negligible $U^{\prime}$ parameter plays an important role in accumulating the predicted $m_W$. However, when $m_{Z^{\prime}}$ increases, the negative $T^{\prime}$ values become harmful in approaching the CDF measured $m_W^{\text{CDF}}$, so the best-fit $\chi^2$ arises swiftly, failing to give a proper fit.

In the $\epsilon_B$ and $\epsilon_W$ parameter space there are two specific combinations of the parameters: $g^{\prime} \epsilon_B = g \epsilon_W$ and $g \epsilon_B = -g^{\prime} \epsilon_W$. The previous one is equivalent to the case that $Z^{\prime}$ only mixes with the photon, and $g^{\prime} \epsilon_B - g \epsilon_W = 0$ results in vanishing $T^{\prime}$ and $U^{\prime}$ according to Eqs.~(\ref{T_Expansion}) and (\ref{DeltaW_Expansion}). On the contrary, when $g \epsilon_B = -g^{\prime} \epsilon_W$, $Z^{\prime}$ only mixes with the SM $Z$ boson, which had been discussed and evaluated in Ref.~\cite{Zeng:2022lkk}, although here we perform a more general analytic calculation.

\begin{figure}[!t]
    \includegraphics[width=0.48\textwidth]{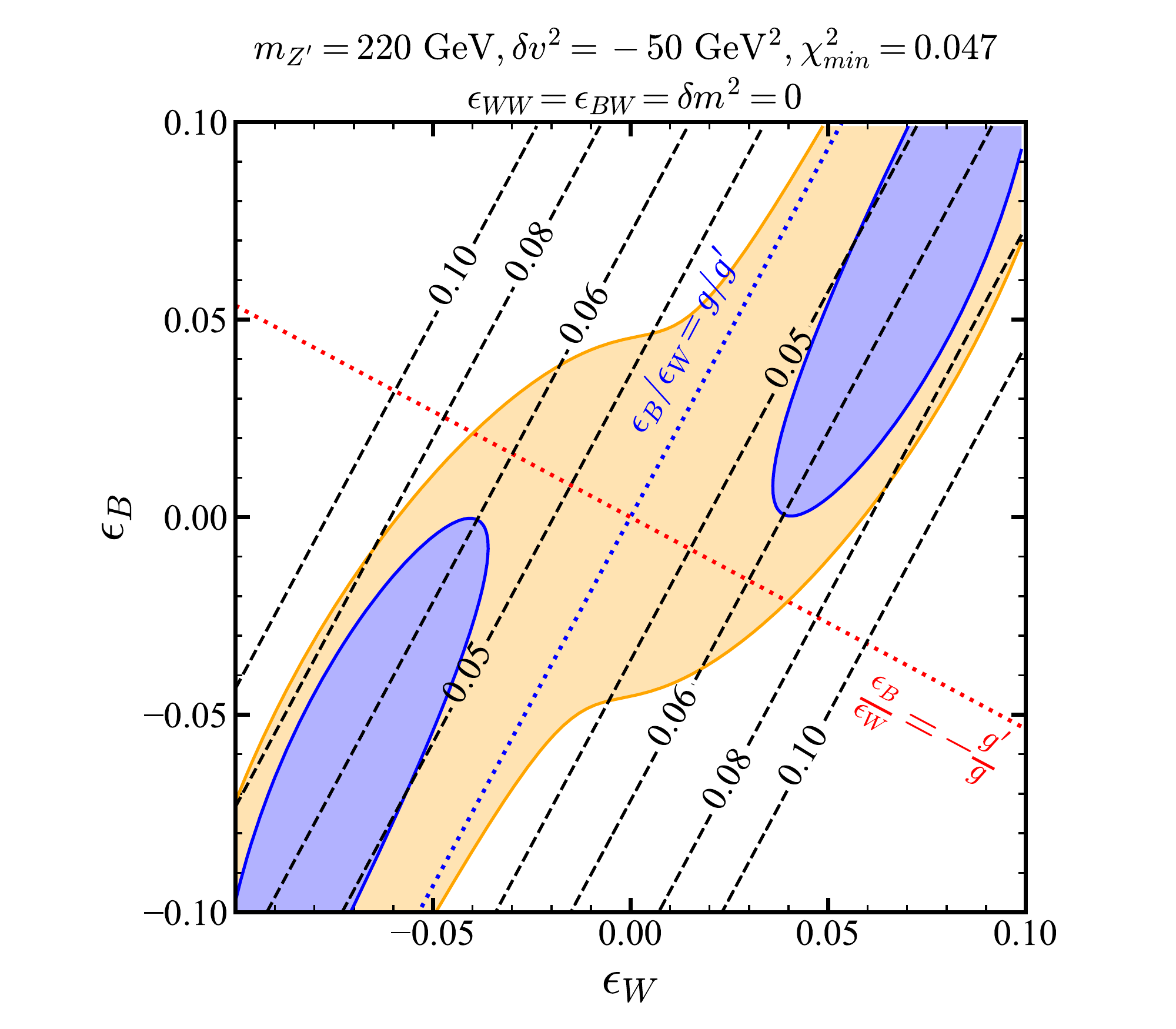}
    \includegraphics[width=0.48\textwidth]{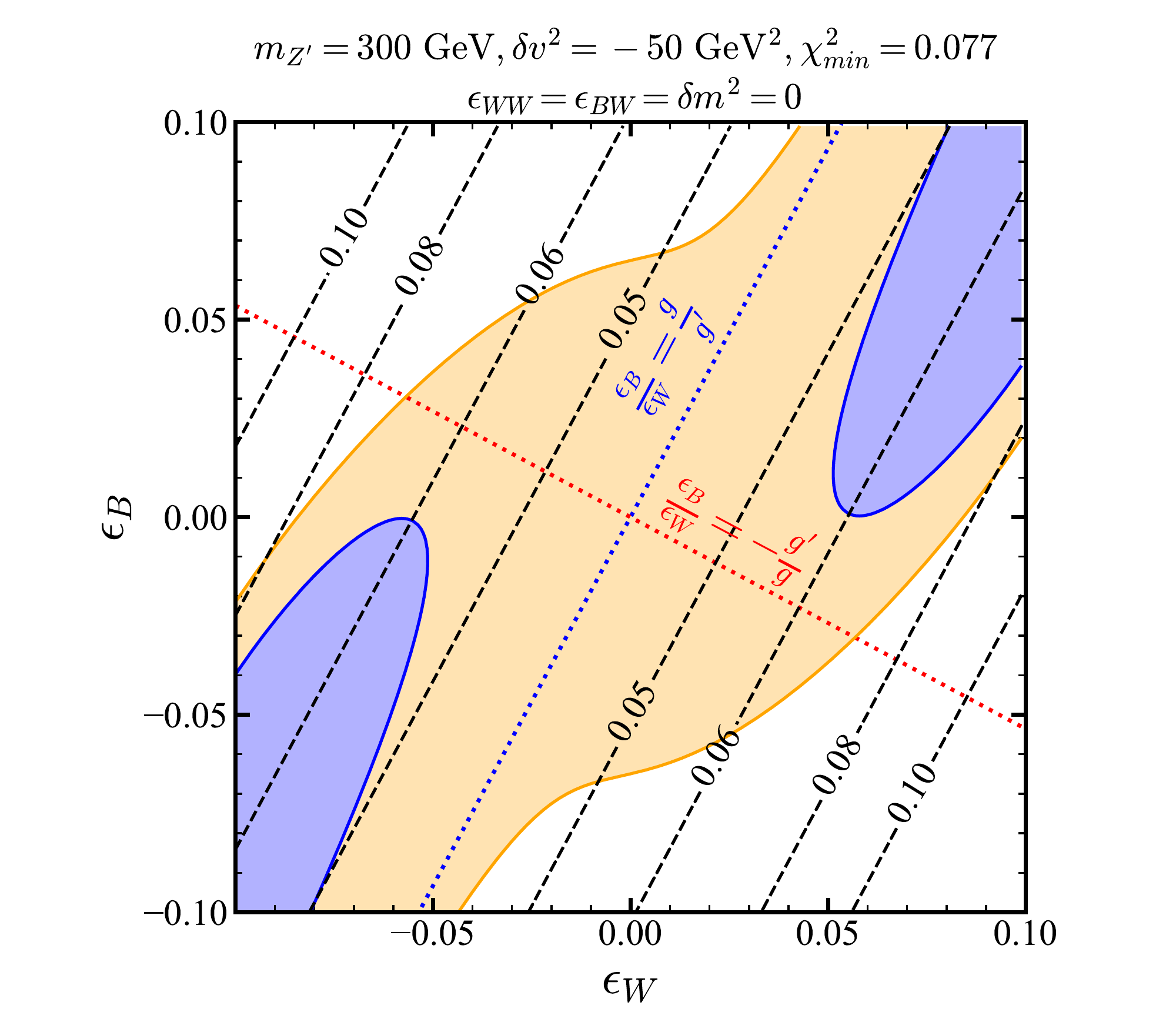}
    \includegraphics[width=0.48\textwidth]{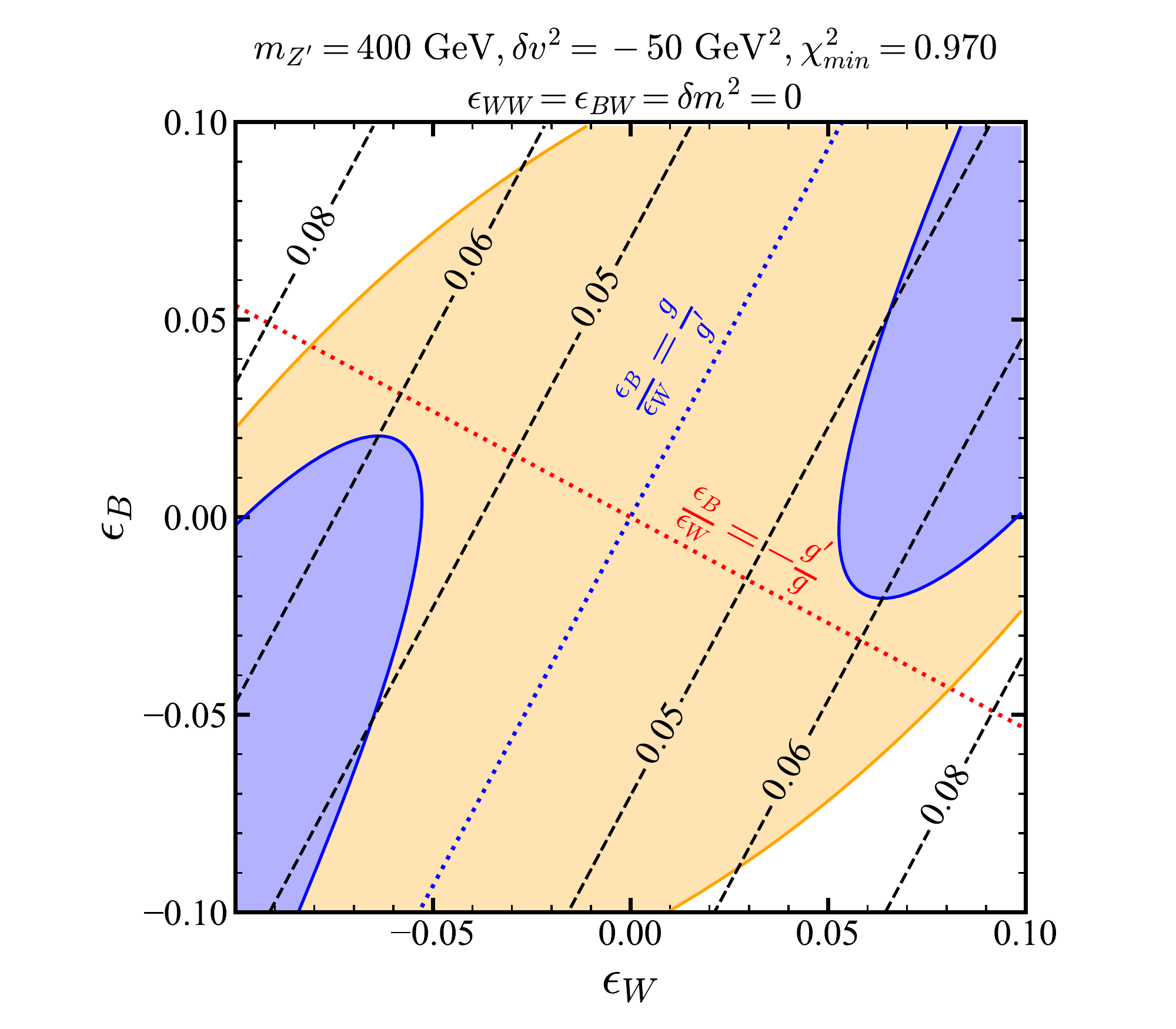}
    \caption{$1\sigma$ (blue) and $2\sigma$ (orange) regions of the fit and $\delta m_W$ contours for $m_{Z^{\prime}}=220~\si{GeV}$ (upper left), $300~\si{GeV}$ (upper right), and $400~\si{GeV}$ (lower) with $\delta v^2=-50~\si{GeV^2}$ and $\epsilon_{WW} = \epsilon_{BW} = \delta m^2 = 0$. The symbols are the same as in Fig.~\ref{mZ_220_NoDeltaV2}.}
    \label{mZ_220_300_400_DeltaV2}
\end{figure}

If we turn on other parameters to contribute to $S^{\prime}$ and $T^{\prime}$, the tension with the global fit can be significantly relieved. Besides the contribution to $S^{\prime}$, the most urgent task is to hoist the value of $T^{\prime}$ from the negative abyss.  As we have mentioned, a positive contribution to $T$ can be realized by adding some extra EW multiplets, e.g., those in the singlet-doublet scalar dark matter model presented in Ref.~\cite{Cai:2017wdu}. From the EFT point of view, the new EW multiplets could generate the $|H^\dag D_\mu H|^2$ operator and hence  $\delta v^2$ at loop level. Both $\delta v^2$ and $\delta m^2$ contribute positively to $T^{\prime}$, while their contributions to $S^{\prime}$ and $U^{\prime}$ are suppressed. Since their impact on $S^{\prime}$, $T^{\prime}$, and $U^{\prime}$ are similar, as an example, we choose to switch on $\delta v^2$ and present the results in Fig.~\ref{mZ_220_300_400_DeltaV2}.

\begin{figure}[!t]
    \includegraphics[width=0.48\textwidth]{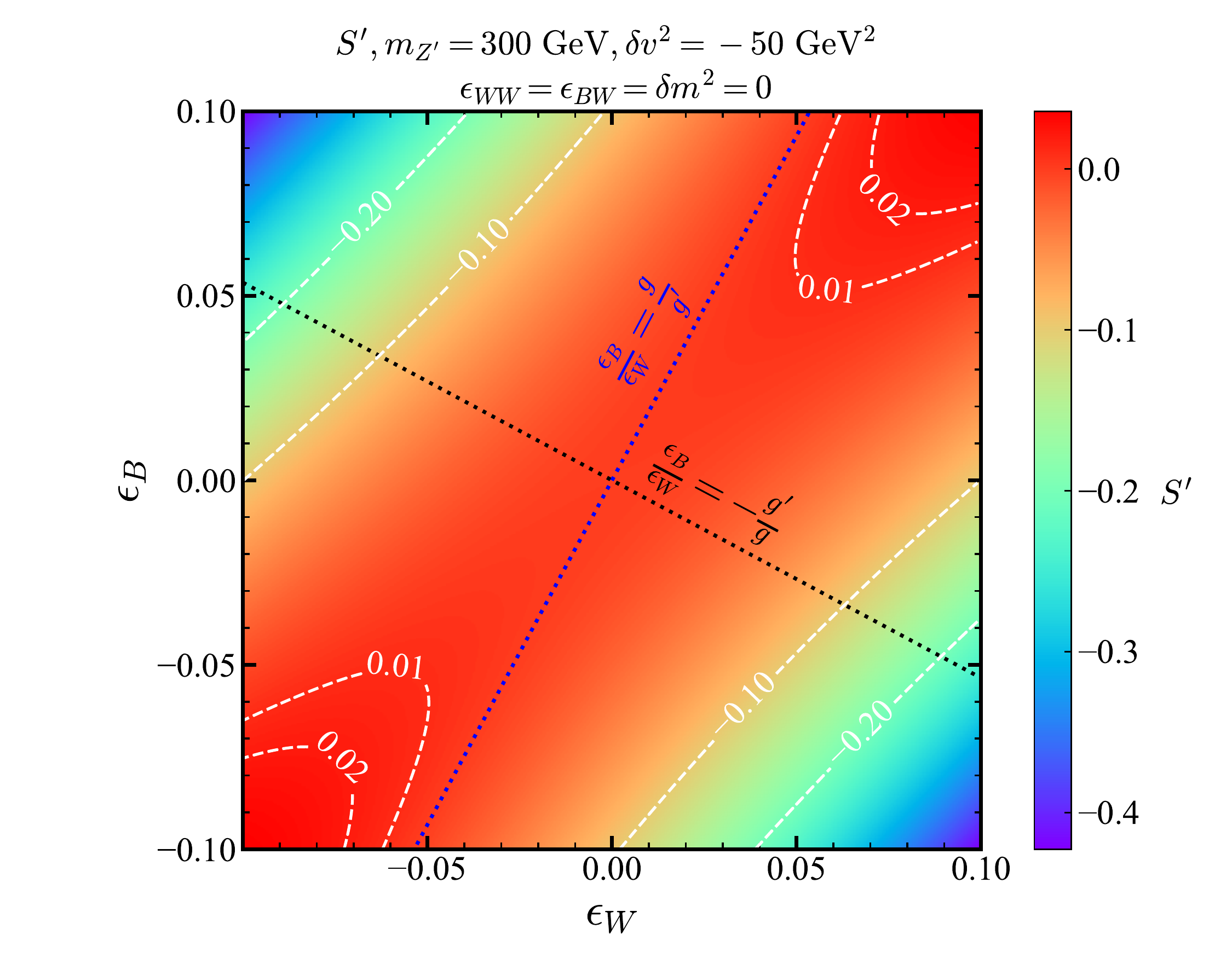}
    \includegraphics[width=0.48\textwidth]{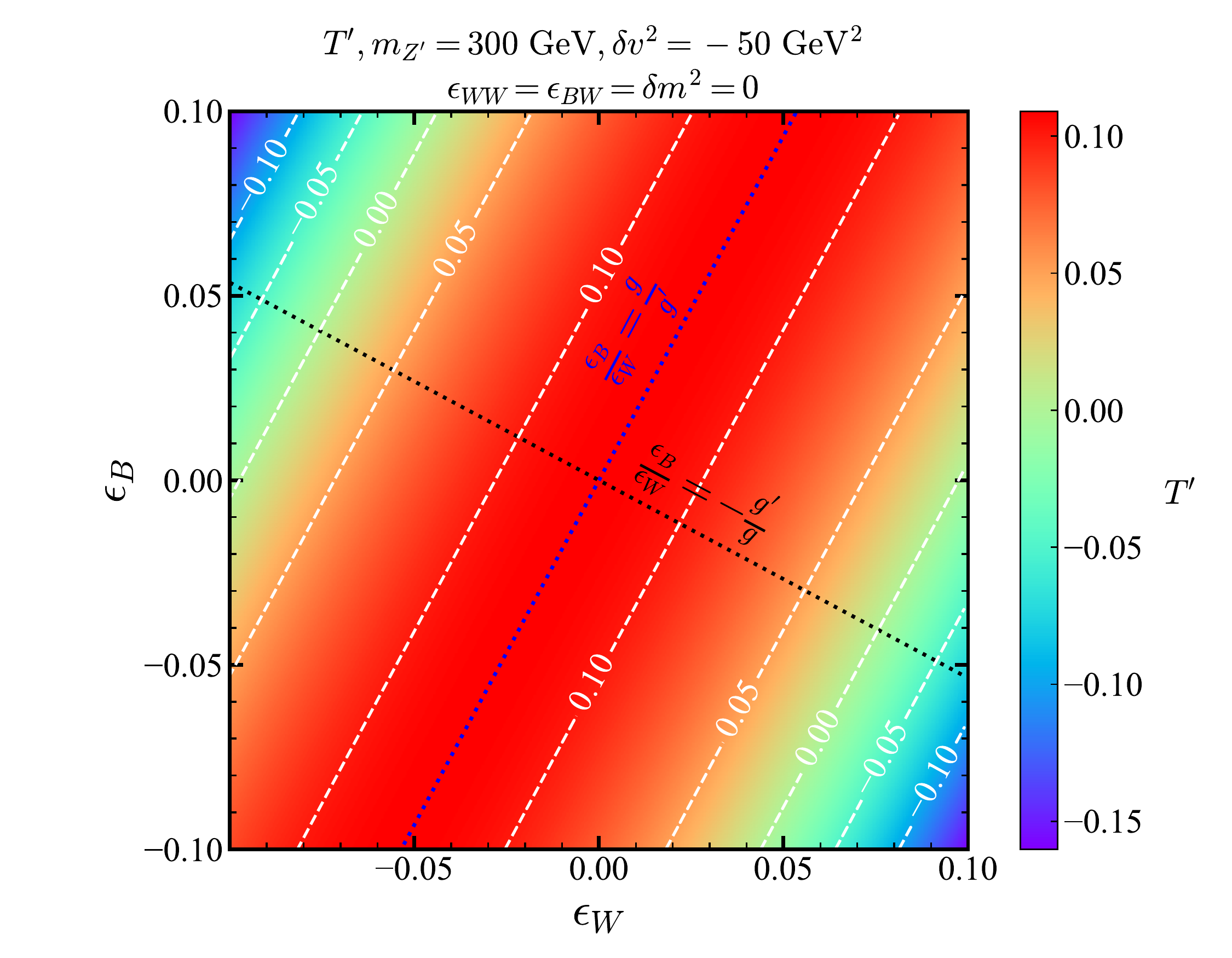}
    \includegraphics[width=0.48\textwidth]{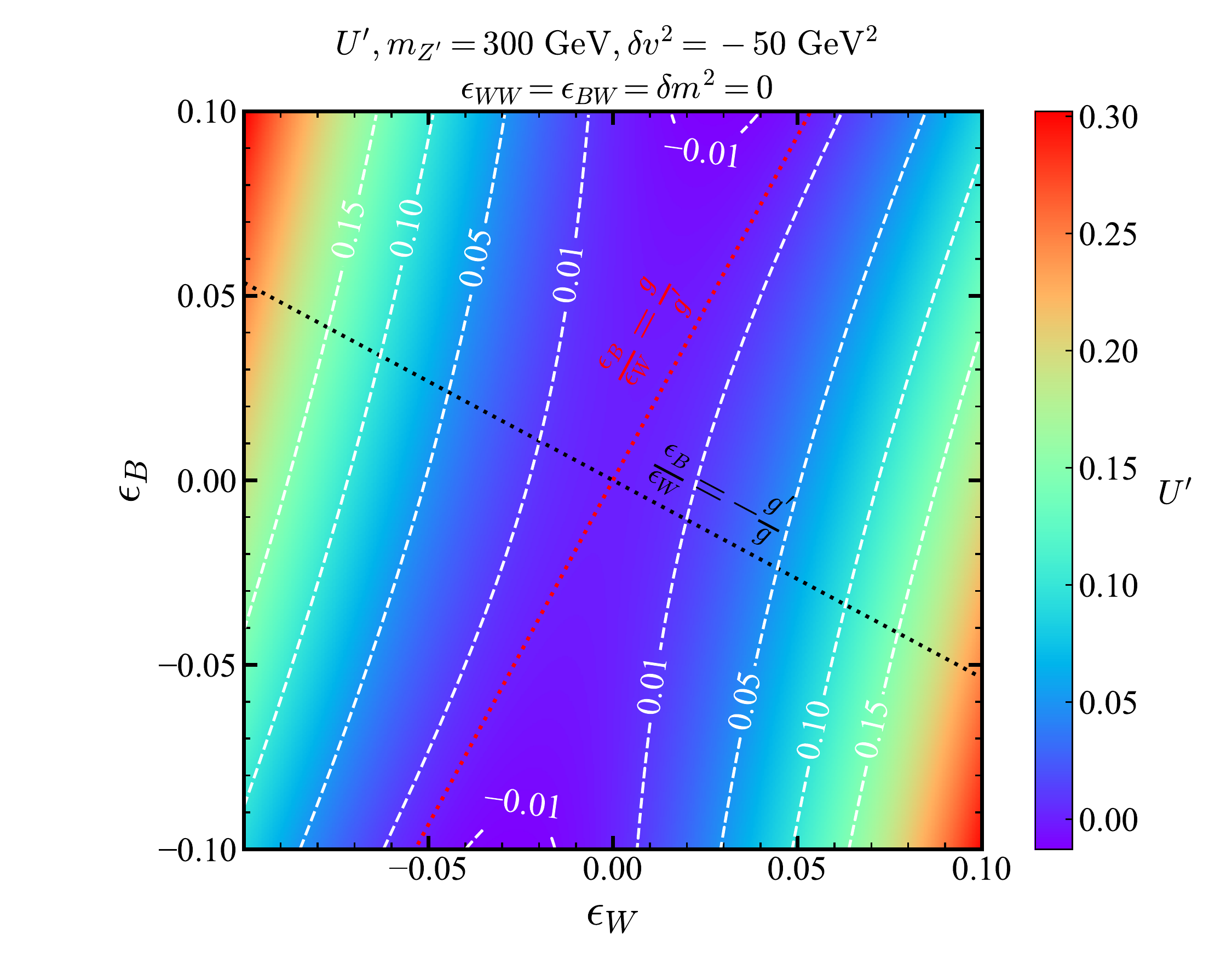}
    \caption{Contours of $S^{\prime}$ (upper left), $T^{\prime}$ (upper right), and  $U^{\prime}$ (lower) for $m_{Z'} = 300~\si{GeV}$, $\delta v^2=-50~\si{GeV^2}$ and $\epsilon_{WW} = \epsilon_{BW} = \delta m^2 = 0$.
    The symbols are the same as in Fig.~\ref{mZ_220_NoDeltaV2}.}
    \label{mZ_300_STU}
\end{figure}

From Fig.~\ref{mZ_220_300_400_DeltaV2} one can easily reckon that the best-fit $\chi^2$ significantly lowers, and thus relatively heavier $Z'$ with $m_{Z^{\prime}} \gtrsim 400$ GeV can also explain the CDF result very well. Comparing the $S^{\prime}$, $T^{\prime}$, and $U^{\prime}$ contours for $m_{Z^{\prime}} = 300$ GeV in Fig.~\ref{mZ_300_STU} with $\delta v^2 = -50~\si{GeV^2}$ and Fig.~\ref{mZ_300_NoDeltaV2} with $\delta v^2 = 0$, we find that a positive $T^{\prime}$ can significantly improve the fit.

\section{Summary and future prospect} \label{Summary}

We have enumerated the possible interactions involving a $Z^{\prime}$ field that induce its mixings with the neutral EW gauge bosons. Both the $\epsilon_B$ and $\epsilon_W$ parameters contribute positively to the non-negligible $U^{\prime}$ parameter, with the price of lowering the $T^{\prime}$ parameter significantly. Appropriate selections of the parameters such as $\delta v^2$ or $\delta m^2$ can accumulate $T^{\prime}$ to relieve the tension between the global fit results and our theoretical predictions. The sufficient increase of the $W$-boson mass can be accomplished to explain the $W$-mass anomaly measured by the CDF II detector within the current collider bounds on the $Z^{\prime}$ boson.

In this paper, we rely on effective field theory with various non-renormalizable operators listed in (\ref{LKin}) and (\ref{LMass}). These operators can arise from charged particles running inside the loops. Building an ultraviolet-complete model inducing all these terms with appropriate coupling strengths will become an important task.

\appendix
\section{Definitions and discussions of the effective $S^{\prime}$, $T^{\prime}$ and $U^{\prime}$}\label{Appendix}

Denote $\Pi_{IJ}(p^2)$ to be the NP contribution to the $g_{\mu\nu}$ coefficient of the vacuum polarization amplitude for EW gauge fields $I$ and $J$, and expand it around $p^2 = 0$,
\begin{equation}
\Pi_{IJ}(p^2) \simeq \Pi_{IJ}(0) + p^2 \Pi'_{IJ}(0) + \frac{(p^2)^2}{2!} \Pi''_{IJ}(0) + \frac{(p^2)^3}{3!} \Pi^{(3)}_{IJ}(0) + \cdots
\end{equation}
Then the Peskin-Takeuchi oblique parameters~\cite{Peskin:1991sw} are defined by
\begin{eqnarray}
\alpha S &=& 4 s_w^2 c_w^2 \left[ \Pi_{ZZ}^{\prime}(0)-\frac{c_w^2-s_w^2}{s_w c_w} \Pi_{ZA}^{\prime}(0) - \Pi_{AA}^{\prime}(0) \right]
= 4 s_w c_w \Pi'_{W^3 B}(0), \nonumber \\
\alpha T &=& \frac{\Pi_{WW}(0)}{m_W^2} - \frac{\Pi_{ZZ}(0)}{m_Z^2}
= \frac{1}{m_W^2} [\Pi_{W^+ W^-}(0) - \Pi_{W^3 W^3}(0)], \nonumber \\
\alpha U &=& 4 s_w^2 \left[ \Pi_{WW}^{\prime}(0) - c_w^2 \Pi_{ZZ}^{\prime}(0) - 2 s_w c_w \Pi_{Z A}^{\prime} ( 0 ) - s_w^2 \Pi_{AA}^{\prime} (0) \right]
\nonumber \\
&=& 4e^2[\Pi'_{W^+ W^-}(0) - \Pi'_{W^3 W^3}(0)]. \label{STUOriginal}
\end{eqnarray}
It is well-known that $S$, $T$, and $U$ correspond to the $H^{\dagger} W^{a}_{\mu \nu} \sigma^a H B^{\mu \nu}$, $H^{\dagger} D_{\mu} H (D^{\mu}H)^{\dagger} H$, and $H^{\dagger} W^a_{\mu \nu} \sigma^a H H^{\dagger} W^{b \mu \nu} \sigma^b H$ operators, respectively. In the literature, their effects on various EW precision observables are evaluated. Comparing the evaluated results with the experimental data, one can constrain the allowed region of $S$, $T$, and $U$.

\begin{figure}
    \centering
    \includegraphics[width=0.3\textwidth]{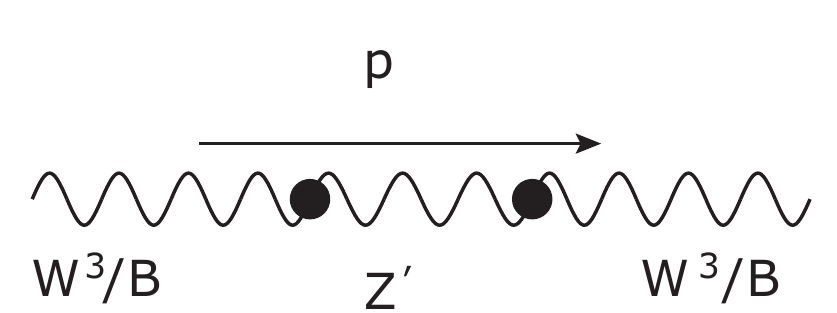}
    \caption{Diagrams integrating out the $Z^{\prime}$ boson to accomodate the effective operators.}
    \label{IntegrateOutZPrime}
\end{figure}

However, other operators which do not contribute to Eq.~(\ref{STUOriginal}) might also shift exactly the same observables to fake the effects of $S$, $T$, and $U$. The $\epsilon_B$ and $\epsilon_B$ terms displayed in the effective Lagrangian (\ref{KineticMixingTerms}) induce the following operators,
\begin{equation}
\mathcal{O}_{BB}^{2i} = B_{\mu \nu} \partial^{2 i} B^{\mu \nu}, \quad
\mathcal{O}_{33}^{2i} = W^3_{ \mu \nu} \partial^{2 i} W^{3 \mu \nu},\quad
\mathcal{O}_{B3}^{2i} = B_{ \mu \nu} \partial^{2 i} W^{3 \mu \nu}, 
\end{equation}
by integrating out the heavy $Z^{\prime}$ in the tree-level $B/W^3$-$Z^{\prime}$-$B/W^3$ oscillation diagrams shown in Fig.~\ref{IntegrateOutZPrime}. For the sake of the completeness of this paper, we also list the following operator,
\begin{eqnarray}
\mathcal{O}_{WW}^{2i} &=& W^{+ \mu \nu} \partial^{2 i} W^{-}_{\mu \nu}. \label{WpWmOperator}
\end{eqnarray}
These set of operators correspond to the higher-order derivatives of the vacuum polarization functions
\begin{eqnarray}
\Pi^{(i)}_{IJ}(p^2)|_{p^2=0} &\sim& \mathcal{O}_{IJ}^{2i},
\end{eqnarray}
where $I,J=B,W^3,W^\pm$. When $i=2$, more oblique parameters $V$, $X$, $Y$, and $W$ can be defined to evaluate the oblique corrections~\cite{Barbieri:2004qk}.
Their definitions are
\begin{eqnarray}
V &=& -\frac{1}{2} m_W^2 [\Pi^{\prime \prime}_{W^3 W^3}(0) - \Pi^{\prime \prime}_{W^+ W^-}(0)],  \nonumber\\
X &=& -\frac{1}{2} m_W^2 \Pi^{\prime \prime}_{W^3 B}(0),  \nonumber\\
Y &=& -\frac{1}{2} m_W^2 \Pi^{\prime \prime}_{B B}(0),  \nonumber\\
W &=& -\frac{1}{2} m_W^2 \Pi^{\prime \prime}_{W^3 W^3}(0). \label{VXYZ}
\end{eqnarray}
In the former studies, the vacuum polarization functions $\Pi_{IJ}(p^2)$ are expanded at most to the second order. However, later we will see, the higher orders account for parts of our results in this paper. Therefore we will give more general discussions below. We should note that alternative definitions of the $S$, $T$, $U$, $V$, $W$, $X$ exist in the literature (See Ref.~\cite{Burgess:1993mg} as an example) which are somehow but incompletely equivalent with the definitions that we adopt from Ref.~\cite{Peskin:1991sw, Barbieri:2004qk}. More straightforward comparisons between the oblique parameters and the effective operators are the advantages of our selections of the oblique parameter definitions.

For brevity of this appendix, let us only preserve the $\epsilon_B$ and $\epsilon_W$ terms, and integrating out the $Z^{\prime}$ boson. To the lowest order, we have
\begin{equation}
\Pi_{BB}(p^2) = \frac{\epsilon_B^2 p^4}{p^2 - m_{Z^{\prime}}^2}, \quad
\Pi_{W^3 W^3}(p^2) = \frac{\epsilon_W^2 p^4}{p^2 - m_{Z^{\prime}}^2}, \quad
\Pi_{W^3 B}(p^2) = \frac{\epsilon_B \epsilon_W p^4}{p^2 - m_{Z^{\prime}}^2}.
\end{equation}
Immediately we obtain
\begin{equation}
\Pi_{BB}^{(i)}(0) = -\frac{i! \epsilon_B^2 }{ m_{Z^{\prime}}^{2i-2}}, \quad
\Pi_{W^3 W^3}^{(i)}(0) = -\frac{i! \epsilon_W^2}{ m_{Z^{\prime}}^{2i-2}}, \quad
\Pi_{W^3 B}^{(i)}(0) =  -\frac{i! \epsilon_B \epsilon_W}{ m_{Z^{\prime}}^{2i-2}},
\end{equation}
for $i \geq 2$, and $\Pi_{IJ}(0)= \Pi_{IJ}'(0) = 0$.

Let us define
\begin{equation}
\alpha_{i} = -\frac{\Pi_{BB}^{(i)}}{i!}, \quad
\beta_{i} = -\frac{\Pi_{W^3 W^3}^{(i)}}{i!}, \quad
\gamma_{i} = -\frac{\Pi_{W^3 B}^{(i)}}{i!},
\end{equation}
and the reciprocal of the resummed propagator (inverse propagator) matrix for $(B,W^3)$ regardless of the tensor part is given by
\begin{eqnarray}
\Pi^{2 \times 2} &=& \begin{pmatrix}
     p^2-\frac{\hat{g}^{\prime 2}}{4} \hat{v}^2-\Pi_{BB}(p^2) &  \frac{\hat{g} \hat{g}^{\prime}}{4} \hat{v}^2 -\Pi_{W^3 B}(p^2)  \\
      \frac{\hat{g} \hat{g}^{\prime}}{4} \hat{v}^2 -\Pi_{W^3 B}(p^2) & p^2- \frac{\hat{g}^{2}}{4} \hat{v}^2 - \Pi_{W^3 W^3}(p^2)
\end{pmatrix}\nonumber \\
&=&
\begin{pmatrix}
     \sum\limits_{i=2}^{\infty} \alpha_i p^{2 i} + p^2 - \frac{\hat{g}^{\prime 2}}{4} \hat{v}^2 & \sum\limits_{i=2}^{\infty} \gamma_i p^{2 i} + \frac{\hat{g} \hat{g}^{\prime}}{4} \hat{v}^2  \\
     \sum\limits_{i=2}^{\infty} \gamma_i p^{2 i} + \frac{\hat{g} \hat{g}^{\prime}}{4} \hat{v}^2 &  \sum\limits_{i=2}^{\infty} \beta_i p^{2 i} + p^2 - \frac{\hat{g}^{2}}{4} \hat{v}^2
\end{pmatrix}.~ \label{Pi44}
\end{eqnarray}

The physical $Z$ and $\gamma$ masses are related to the solution of the equation $\det(\Pi_{4 \times 4})=0$. Similar to the $\alpha_i=0$, $\beta_i=0$, $\gamma_i=0$ $(i\geq 2)$ case, we can turn to the mass eigenstates through the EW rotation matrix
\begin{equation}
V_{\text{SM}}^{2 \times 2} = \begin{pmatrix}
-\frac{\hat{g}^{\prime}}{\sqrt{\hat{g}^{\prime 2} + \hat{g}^2}} & \frac{\hat{g}}{\sqrt{\hat{g}^{\prime 2} + \hat{g}^2}} \\
\frac{\hat{g}}{\sqrt{\hat{g}^{\prime 2} + \hat{g}^2}} & \frac{\hat{g}^{\prime}}{\sqrt{\hat{g}^{\prime 2} + \hat{g}^2}}
\end{pmatrix},
\end{equation}
so that
\begin{eqnarray}
(V_{\text{SM}}^{2 \times 2})^\mathrm{T} \Pi^{2 \times 2} V_{\text{SM}}^{2 \times 2} =
\begin{pmatrix}
     \sum\limits_{i=2}^{\infty} a_i p^{2 i} + p^2 - \hat{m}_Z^2 & \sum\limits_{i=2}^{\infty} c_i p^{2 i}  \\
     \sum\limits_{i=2}^{\infty} c_i p^{2 i}  &  \sum\limits_{i=2}^{\infty} b_i p^{2 i} + p^2 
\end{pmatrix}, \label{Pi44_PreDiagonalized}
\end{eqnarray}
where $\hat{m}_Z^2 = (\hat{g}^2 + \hat{g}^{\prime 2})\hat{v}^2/4$, and
\begin{eqnarray}
\begin{pmatrix}
     a_i & c_i \\
     c_i & b_i
\end{pmatrix} = (V_{\text{SM}}^{2 \times 2})^\mathrm{T} \begin{pmatrix}
     \alpha_i & \gamma_i \\
     \gamma_i & \beta_i
\end{pmatrix} V_{\text{SM}}^{2 \times 2}.
\end{eqnarray}
Notice that there is a massless solution $p^2=0$ with the eigenvector $(0, 1)^\mathrm{T}$ corresponding to the massless photon.

Another eigenvector can be parameterized as $(1, t)^\mathrm{T}$, so we have to solve the equation
\begin{eqnarray}
\begin{pmatrix}
     \sum\limits_{i=2}^{\infty} a_i p^{2 i} + p^2 - \hat{m}_Z^2 & \sum\limits_{i=2}^{\infty} c_i p^{2 i}  \\
     \sum\limits_{i=2}^{\infty} c_i p^{2 i}  &  \sum\limits_{i=2}^{\infty} b_i p^{2 i} + p^2 
\end{pmatrix} 
\begin{pmatrix}
     1  \\
     t 
\end{pmatrix} = 0. \label{MasterEquation}
\end{eqnarray}
When all $a_i,b_i,c_i=0$, we derive $t=0$. If $\alpha_i m_Z^{2 i -2}$, $\beta_i m_Z^{2 i -2}$, and $\gamma_i m_Z^{2 i -2}$ (or $a_i m_Z^{2 i -2}$, $b_i m_Z^{2 i -2}$, and $c_i m_Z^{2 i -2}$) are considered to be of the same order which are much smaller than $1$ regardless of the power index, one can expect that $t \ll 1$ is also of the same order. Then we can solve Eq.~(\ref{MasterEquation}) perturbatively by discarding all the higher-order terms,
\begin{eqnarray}
\left\lbrace \begin{array}{l}
  \sum\limits_{i=2}^{\infty} a_i p^{2 i} + t \sum\limits_{i=2}^{\infty} c_i p^{2 i} + p^2 - \hat{m}_Z^2 = 0, \\
  \sum\limits_{i=2}^{\infty} c_i p^{2 i} + t p^2=0,
\end{array} \right.  \Rightarrow \left\lbrace \begin{array}{l}
   p^2 \simeq \hat{m}_Z^2 - \sum\limits_{i=2}^{\infty} a_i \hat{m}_Z^{2 i}, \\
   t \simeq -\sum\limits_{i=2}^{\infty} c_i m_Z^{2 i-2}.
\end{array} \right.  \label{Solution}
\end{eqnarray}
After rotating $\Pi^{2 \times 2}$ with
\begin{eqnarray}
V_t = \begin{pmatrix}
     1 & t  \\
     0 & 1
\end{pmatrix}, \label{tTransform}
\end{eqnarray}
we acquire the ``diagonalized'' inverse propagator
\begin{eqnarray}
\Pi^{2 \times 2}_{\text{M}} = V_t^\mathrm{T} (V_{\text{SM}}^{2 \times 2})^\mathrm{T} \Pi^{2 \times 2} V_{\text{SM}}^{2 \times 2} V_t. \label{Pi44_Diagonalized}
\end{eqnarray}

The element $\Pi^{2 \times 2}_{\text{M}, 00}(p^2) \simeq \sum\limits_{i=2}^{\infty} a_i p^{2 i} + \sum\limits_{i=2}^{\infty} c_i p^{2 i} + p^2 - \hat{m}_Z^2$ is the inverse propagator of the physical $Z$ boson near its pole, and the reciprocal of the residue of the pole is
\begin{eqnarray}
\left. \frac{\partial \Pi^{2 \times 2}_{\text{M}, 00}(p^2)}{\partial p^2} \right|_{p^2 = \hat{m}_Z^2 - \sum\limits_{i=2}^{\infty} a_i \hat{m}_Z^{2 i}} \simeq 1 + \sum\limits_{i=2}^{\infty} i a_i p^{2 i-2}. \label{Normalization}
\end{eqnarray}
A complete calculation of the NC and CC terms requires normalizing out this factor. Therefore, comparing (\ref{Solution}), (\ref{tTransform}) and (\ref{Normalization}) with the corresponding terms of Eq.~(5) and (6) in Ref.~\cite{Burgess:1993vc}, we can see that the physical mass shift $- \sum\limits_{i=2}^{\infty} a_i \hat{m}_Z^{2 i}$, the rotation parameter $t$, and the field normalization factor $\sum\limits_{i=2}^{\infty} i a_i p^{2 i-2} $ are equivalent to $(z-C) \hat{m}_Z^2$, $G$, and $C$ defined in Ref.~\cite{Burgess:1993vc}, respectively. Straightforwardly casting the symbols there, we have
\begin{equation}
(z-C) = -\sum\limits_{i=2}^{\infty} a_i \hat{m}_Z^{2 i -2},\quad
G = -\sum\limits_{i=2}^{\infty} c_i \hat{m}_Z^{2 i-2},\quad
C = \sum\limits_{i=2}^{\infty} i a_i \hat{m}_Z^{2 i-2}.
\end{equation}

If the operator (\ref{WpWmOperator}) arises, the inverse propagator of the $W$-boson regardless of the tensor part is given by
\begin{eqnarray}
p^2-\hat{m}_W^2-\Pi_{W^+ W^-}(p^2) = \sum\limits_{i=2}^{\infty} d_i p^{2 i} + p^2 - \hat{m}_W^2, \label{WFullPropagator}
\end{eqnarray}
where $\hat{m}_W = \hat{g} v/2$. Again, solving $p^2-\hat{m}_W^2-\Pi_{W^+ W^-}(p^2) = 0$ gives the solution of the physical $W$-boson mass,
\begin{eqnarray}
m_W^2 = p^2 = \hat{m}_W^2 - \sum\limits_{i=2}^{\infty} d_i \hat{m}_W^{2 i}.
\end{eqnarray}
The reciprocal of the residue of the propagator near the pole accordingly becomes
\begin{eqnarray}
1-\left. \frac{\partial \Pi_{W^+ W^-}(p^2)}{\partial p^2} \right|_{p^2 = \hat{m}_W^2 - \sum\limits_{i=2}^{\infty} d_i \hat{m}_Z^{2 i}} \simeq 1 + \sum\limits_{i=2}^{\infty} i d_i p^{2 i-2}.
\end{eqnarray}
Comparing with the corresponding terms of Eq.~(4) and (6) in Ref.~\cite{Burgess:1993vc} and casting its symbols again similarly, we obtain
\begin{equation}
(w-B) = -\sum\limits_{i=2}^{\infty} d_i \hat{m}_W^{2 i -2},\quad
B = \sum\limits_{i=2}^{\infty} i d_i \hat{m}_W^{2 i-2}.
\end{equation}

Therefore, if we follow Eqs.~\eqref{SParameter}, \eqref{TParameter} and \eqref{WMass2U} to compute the oblique paramters, we are actually substituting our effective $B$, $C$, $G$, $w$, and $z$ into Eq.~(2) in Ref.~\cite{Burgess:1993vc} to acquire the effective $S^{\prime}$, $T^{\prime}$, and $U^{\prime}$ values. Before displaying the results, we should note that the $w$ parameter appeared in Eq.~(15) of Ref.~\cite{Burgess:1993vc} should be abolished in our case, since $G_F$ is defined in the low-$p^2$ limit so that all the $d_i p^2$ terms in Eq.~\eqref{WFullPropagator} become ineffective. Therefore, all the $w$ symbols corresponding to Eqs.~(17) and (18) in Ref.~\cite{Burgess:1993vc} should be discarded. Our $S'$ and $T'$ in Eqs.~\eqref{SParameter} and \eqref{TParameter} are derived by matching Eq.~(23) in Ref.~\cite{Burgess:1993vc}, and $w$ there comes from Eq.~(17), so it completely disappears. We also utilized Eq.~(20) in Ref.~\cite{Burgess:1993vc} to accommodate our definition of $U'$, and it is easily realized that $w$ within $A-C-w+z$ disappears again, and the remaining $w$ there was absorbed by $U'$. Therefore, Eq.~(2) in Ref.~\cite{Burgess:1993vc} should be adjusted to
\begin{eqnarray}
\alpha S^{\prime} &=& 4 s_w^2 c_w^2 \left(-C-\frac{c_w^2-s_w^2}{c_w s_w} G\right), \nonumber \\
\alpha T^{\prime} &=& -z, \nonumber \\
\alpha U^{\prime} &=& 4 s_w^4 \left[ -\frac{1}{s_w^2} (B-w) + \frac{c_w^2}{s_w^2} C - \frac{2 c_w}{s_w} G \right]. \label{SpTpUp1}
\end{eqnarray}

If $S$, $T$, and $U$ defined in Eq.~(\ref{STUOriginal}) also exists, they will also contribute to $S^{\prime}$, $T^{\prime}$ and $U^{\prime}$, and Eq.~(\ref{SpTpUp1}) then becomes
\begin{eqnarray}
\alpha S^{\prime} &=& \alpha S + 4 s_w^2 c_w^2 \left(-C-\frac{c_w^2-s_w^2}{c_w s_w} G\right), \nonumber \\
\alpha T^{\prime} &=& \alpha T  -z, \nonumber \\
\alpha U^{\prime} &=& \alpha U + 4 s_w^4 \left[ -\frac{1}{s_w^2} (B-w) + \frac{c_w^2}{s_w^2} C - \frac{2c_w}{s_w} G \right].
\end{eqnarray}
Expressing with $\alpha_i$, $\beta_i$, $\gamma_i$, and $d_i$, we arrive at
\begin{eqnarray}
\alpha S^{\prime} - \alpha S
&=&  \sum\limits_{i=2}^{\infty} 4 s_w^2 c_w^2 m_Z^{2 i-2} \bigg\lbrace [-(i-1) s_w^2 - c_w^2] \alpha_i + [-(i-1) c_w^2 - s_w^2] \beta_i
\nonumber \\
&&\hspace{7.4em} + \frac{s_w^4 + c_w^4 + (2 i-2) s_w^2 c_w^2 }{s_w c_w} \gamma_i \bigg\rbrace, \nonumber \\
 \alpha T^{\prime} - \alpha T &=& -\sum\limits_{i=2}^{\infty} (i-1) (\alpha_i s_w^2 - 2 \gamma_i s_w c_w + \beta_i c_w^2) m_Z^{2 i-2}, \nonumber \\
 \alpha U^{\prime} - \alpha U &=& -\sum\limits_{i=2}^{\infty} \frac{(i-1) d_{i} m_W^{2 i-2}}{s_w^2} + \sum\limits_{i=2}^{\infty} \frac{i c_w^2}{s_w^2} m_Z^{2 i-2} ( \alpha_i s_w^2 - 2 \gamma_i s_w c_w + \beta_i c_w^2) \nonumber \\
&& +  \sum\limits_{i=2}^{\infty} \frac{2 c_w}{s_w}m_Z^{2 i-2} [\gamma_i (c_w^2 - s_w^2) + \beta_i s_w c_w - \alpha_i s_w c_w]. 
\end{eqnarray}

If we only preserve the $i=2$ terms, with the oblique parameters defined in Eq.~(\ref{VXYZ}), we derive
\begin{eqnarray}\label{STUprime}
\alpha S^{\prime} &\simeq& \alpha S + 4 s_w^2 \left(-Y -W  + \frac{s_w}{c_w} X \right), \nonumber \\
\alpha T^{\prime} &\simeq& \alpha T - \frac{s_w^2}{c_w^2} Y - W + \frac{2 s_w}{c_w} X, \nonumber \\
\alpha U^{\prime} &\simeq& \alpha U - 4 s_w^2 \left( -W-V+\frac{2s_w}{c_w} X \right),
\end{eqnarray}
which is exactly compatible with the expressions of $\varepsilon_1$, $\varepsilon_2$, and $\varepsilon_3$ in Ref.~\cite{Barbieri:2004qk}. These $\varepsilon_i$ parameters had originally been suggested in Ref.~\cite{Altarelli:1990zd}, where they are considered to be equivalent to the $S$ and $T$ parameters since the contributions from higher derivatives of the gauge fields are neglected. Ref.~\cite{Barbieri:2004qk} includes the influence from $V$, $W$, $X$, and $Y$ without giving the detailed derivations, which are contributions at the $p^4$ order. Besides these, below we will show that the $p^6$ order also arises. 

When only the kinetic mixings between $Z^\prime$ and the EW gauge bosons are considered, the oblique parameters are given by
\begin{eqnarray}\label{STUVWYX}
S=T=U=0,\quad
V=W=\frac{\epsilon_W^2m_W^2}{m_{Z'}^2},\quad
Y=\frac{\epsilon_B^2m_W^2}{m_{Z'}^2},\quad
X=\frac{\epsilon_B\epsilon_Wm_W^2}{m_{Z'}^2}.
\end{eqnarray}
Note that $X$ and $V$ arising from dim-8 and dim-10 SMEFT operators are usually expected to be negligible, but they have the same order of magnitude as $W$ and $Y$ in our model, since they are generated by the mediation of a $Z^\prime$ boson which is not much heavier than the EW scale.
Substituting Eq.~\eqref{STUVWYX} into Eq.~\eqref{STUprime}, we derive
\begin{eqnarray}
\alpha S^{\prime} &\simeq&  -\frac{g^2g'^2v^2(\epsilon_B^2+\epsilon_W^2)}{(g^2+g'^2)m_{Z'}^2}+\frac{gg'v^2\epsilon_B\epsilon_W}{m_{Z'}^2}, \nonumber \\
\alpha T^{\prime} &\simeq& -\frac{(g'\epsilon_B-g\epsilon_W)^2}{4m_{Z'}^2}, \nonumber \\
\alpha U^{\prime} &\simeq& \frac{2g'^2v^2(g^2\epsilon_W^2-gg'\epsilon_B\epsilon_W)}{(g^2+g'^2)m_{Z'}^2}.
\end{eqnarray}
These reproduce $S^\prime$ and $T^\prime$ given by expanding Eq.~\eqref{S_Expansion} and \eqref{T_Expansion} to the $m_{Z'}^{-2}$ order. 
Note that there is no $\epsilon_B^2$ term in $U'$ if we only include the second order contribution, $\Pi^{(2)}_{BB}(p^2)$ ($Y$ parameter). The leading $\epsilon_B^2$ contribution to $U'$ comes from $\Pi^{(3)}_{BB}(0)$ (coefficient of $p^6$ term) and yields $\alpha\delta U^\prime= 4s_w^4c_w^2\epsilon_B^2m_Z^4/m_{Z'}^4$, which reproduces the result given in Ref.~\cite{Holdom:1990xp}.

\begin{acknowledgements}
We thank to Yu-Pan Zeng for helpful discussions and communications. This work is supported in part by the National Natural Science Foundation of China under Grants Nos. 12005312, 11875327, 11905300, and 11805288, the Fundamental Research Funds for the Central
Universities, the Natural Science Foundation of Guangdong Province, and the Sun Yat-Sen University Science Foundation. 
\end{acknowledgements}

\bibliographystyle{utphys}
\bibliography{ref}

\end{document}